\documentclass[manuscript]{acmart}
\usepackage{multirow} 
\usepackage{array} 
\usepackage[table,xcdraw]{xcolor}

\AtBeginDocument{%
  }

\setcopyright{acmlicensed}
\copyrightyear{2026}
\acmYear{2026}
\acmDOI{XXXXXXX.XXXXXXX}

\acmConference[CSCW '26]{The ACM Conference on Computer-Supported Cooperative Work and Social Computing}{June 03--05,
  2026}{Salt Lake City, UT}
\acmISBN{978-1-4503-XXXX-X/18/06}

\begin{document}

\title{PRISM: Evaluating a Rule-Based, Scenario-Driven Social Media Privacy Education Program for Young Autistic Adults}



\author{Kirsten Chapman}
\affiliation{%
  \institution{Brigham Young University}
  \country{U.S.A}
}
\author{Garrett Smith}
\affiliation{%
  \institution{Brigham Young University}
  \country{U.S.A}
}
\author{Kaitlyn Klabacka}
\affiliation{%
  \institution{Brigham Young University}
  \country{U.S.A}
}
\author{Joseph Thomas Bills}
\affiliation{%
  \institution{Brigham Young University}
  \country{U.S.A}
}
\author{Addisyn Bushman}
\affiliation{%
  \institution{Brigham Young University}
  \country{U.S.A}
}
\author{Terisa Gabrielsen}
\affiliation{%
  \institution{Brigham Young University}
  \country{U.S.A}
}
\author{Pamela J. Wisniewski}
\affiliation{%
  \institution{International Computer Science Institute}
  \country{U.S.A}
}
\author{Xinru Page}
\affiliation{%
  \institution{Brigham Young University}
  \country{U.S.A}
}

\renewcommand{\shortauthors}{Chapman et al.}

\newcommand{\oldedit}[1]{{\color{black} #1}}
\newcommand{\edit}[1]{{\color{black} #1}}
\newcommand{\finaledit}[1]{{\color{black} #1}}

\begin{abstract}
Young autistic adults may garner benefits through social media but also disproportionately experience privacy harms. Prior research found that these harms often stem from perceiving the affordances of social media differently than the general population, leading to \edit{unintentional} risky behaviors and interactions with others. While educational interventions have been shown to increase social media privacy literacy for the general population, research has yet to focus on effective educational interventions for autistic young adults. We address this gap by developing and deploying \edit{\textbf{\textit{Privacy Rules for Inclusive Social Media}} (PRISM)}, a classroom-based educational intervention tailored to the unique risks and neurodevelopmental differences of this population. Twenty-nine autistic students with substantial (level 2) support needs participated in a 14-week social media privacy literacy class. During these classes, participants often communicated their existing rule-based ``all or nothing'' approaches to privacy management (such as completely disengaging from social media to avoid privacy issues). Our course focused on empowering them \edit{by providing} more nuanced \edit{guidance on} safe privacy practices through the use of \edit{scenario-based formats} and contextual, rule-based scenarios. Using pre- and post-knowledge assessments for each of our 6 course topics, our intervention led to a statistically significant increase in their \finaledit{making} safer social media privacy decisions. We conclude with recommendations for how privacy educators and technology designers can leverage \edit{neuro-affirming} educational interventions to increase privacy literacy for autistic social media users.
\end{abstract}

\received{13 May 2025}

\maketitle

\section{Introduction}
Social media use can increase social capital and well-being \cite{ellison2015social, ellison2007benefits, burke2010social, burke2016once, burke2011social, ellison2014cultivating} and has been shown to be especially beneficial for helping autistic young adults (YAs)\footnote{There are wide variations in preference for person-first vs identify-first language. We use identify-first language in this paper based on the preferences of our participants.} \edit{to overcome social and other challenges by supporting their day-to-day and developmental needs} \cite{page2022perceiving, wang2020benefits, gillespie2021using, mazurek2013social}. \edit{For instance,} connecting online empowers many autistic YAs to \edit{solve} transportation and mobility challenges \cite{lubin2016transportation}, \edit{as well as giving them access to job seeking opportunities and resources that might not otherwise have been available} \cite{mowbray2021using, lu2021uncovering}. However, autistic YAs are also more likely to experience \edit{disproportionate} privacy violations (and resulting harms) when using social media \cite{page2022perceiving, wang2020benefits}. Prior research has \edit{found} increased cyberbullying, physical harms, and decreased mental health correlated with social media use \edit{for autistic users} and points to how many privacy violations occur because autistic YAs perceive the affordances of social media differently than the general population \cite{page2022perceiving, triantafyllopoulou2022social}. In other words, they often interpret the meaning of user interface elements differently. For example, autistic YAs may interpret a Facebook ``friend'' as someone trustworthy and a true friend in every sense of the word, accepting many friend requests in hopes of making friends. This leads to connecting with bad actors who gain access to them and their private information, enabling financial, emotional, or physical harm. \edit{Subsequent HCI research \cite{cullen2024towards}} found that \edit{supporting adults} (e.g., parents, case workers) \edit{of autistic YAs with significant support needs often had to mediate social media use to help them navigate safety issues. This research} uncovered how self-regulation mechanisms \edit{of autistic YAs, even acknowledged by the YAs themselves,} were \edit{at times} inadequate, establishing the need for increased social media privacy and safety literacy \edit{for this population.} However, more work needs to be done to understand the various self-regulation techniques used for a range of common social media privacy risks \edit{for autistic YAs who have significant support needs.} \edit{The CSCW community has a tradition of examining important topics like raising privacy awareness \cite{wang2013online, zhang2016privacy, jia2015risk}, education regarding social computing \cite{sun2017enhancing, bailey2025students, davis2018evaluating}, and supporting vulnerable populations \cite{van2023understanding, tang2025beyond, li2025challenges}. This paper contributes to these streams of CSCW research by supporting safer social media use for a population that is disproportionately exposed to privacy risks through the design and evaluation of a targeted  educational intervention in the classroom.}

Researchers in \edit{the CSCW and broader HCI} community have developed social media privacy education focused on \edit{vulnerable populations, such as} adolescents \cite{khan2024teaching}, university students \cite{egelman2016teaching}, and older adults \cite{aly2024tailoring}. However, \edit{this body of work has not created nor empirically evaluated privacy-literacy interventions designed specifically for autistic social media users, whose developmental needs and communication styles differ substantially from those of these prior audiences \cite{iovannone2003effective}. As a result, it remains unclear how autistic young adults (YAs) currently understand and manage privacy risks on social media, or whether existing strategies align with their lived experiences and support needs. Evidence-based approaches such as rule-governed behavior, which emphasize explicit instruction over implicit, trial-and-error learning, have been shown to promote independence for autistic individuals in offline contexts \cite{bradley2022rule}. However, these approaches have not yet been applied or studied in the context of social media privacy education for autistic YAs. This gap raises open questions about how privacy education should be designed \textit{\textbf{for young autistic adults with substantial support needs}}, and which instructional features best support learning and engagement for this population. Consequently, existing privacy-education approaches may be ill-suited to help autistic users recognize and avoid risky social media behaviors. To address these gaps,} we designed and \edit{evaluated} a social media privacy educational intervention \edit{that accounts for autistic communication styles and preferences for rule-based instruction. We ask the following research questions in regards to this educational intervention and target population:} 

\begin{quote}
    \textbf{RQ1:} \textit{What existing privacy management strategies \edit{and underlying beliefs} do autistic YAs \edit{exhibit} when using social media \edit{that may unintentionally constrain their overall experience?}} \\
    \textbf{RQ2:} \textit{\edit{How can rule-based, scenario-driven social media privacy education be developed to effectively improve privacy decision-making learning outcomes to empower autistic YAs?}}\\
    \oldedit{\textbf{RQ3:} \textit{How do participant's \edit{classroom participation patterns (e.g., verbal contribution, peer interaction) relate to variation in learning outcomes?}}}
\end{quote}

To answer these research questions, \edit{we introduce PRISM, a novel educational program to support the development of social media privacy and trust awareness skills among autistic young adults} for Facebook and Instagram (platforms commonly used by autistic YAs \cite{page2022perceiving}). The intervention consisted of weekly 50-minute classes over the course of 14 weeks. We administered the intervention to students in a residential program for transition-aged autistic YAs (18-30's) \edit{with significant support needs}. The course topics were drawn from prior privacy education curricula \cite{egelman2016teaching, hardin2020digital, khan2024teaching} augmented with privacy-related problems specific to this population identified in prior work \cite{page2022perceiving, cullen2024towards}. \edit{In contrast to prior social media literacy interventions, it uses rule-based instruction which caters to autistic learning styles. This paper reports on the design and structure of the PRISM program and presents an analysis of qualitative and quantitative learning outcomes, data gathered from recordings of PRISM sessions, and an end-of-program evaluation.}


\edit{The primary contribution of this work is the design, deployment, and empirical evaluation of PRISM, a classroom-based social media privacy education program tailored for young autistic adults with substantial support needs. Through pre–post assessments across 6 instructional modules, we demonstrate the efficacy of rule-based, scenario-driven instruction in improving social media privacy decision-making for this population (RQ2). In addition, by analyzing in-situ classroom discussions regarding students' current social media privacy practices and attitudes, this work provides supporting qualitative insights into autistic young adults’ existing privacy beliefs and rule-based strategies, which help contextualize how and why such an intervention can be effective (RQ1). Further, by examining classroom participation, we provide exploratory evidence that differences in participants’ collaborative behaviors and levels of classroom engagement were associated with variation in learning outcomes (RQ3). Namely, those who did not have a positive learning outcome did not exhibit certain types of engagement. Overall, these contributions address a gap in social media privacy education for autistic young adults by offering and evaluating a neuro-affirming, evidence-based curriculum, and by identifying implications for the design of privacy education and social technologies that better support neurodiverse users.}

\section{Background and Related Work}
In this section, we summarize previous research related to our study. \edit{First, we explore the experiences of autistic individuals with social technologies, including platforms developed specifically for this population. Then,} we discuss the social media and privacy \edit{harms experienced by autistic individuals}. Next, we examine existing privacy education research. Finally, we delve into the learning needs of autistic individuals in education. 

\subsection{\edit{Social Media and Autistic Users}}
Social media has become a ubiquitous communication tool in the United States \cite{pew2024socialmedia}, \edit{with 79.6\% of autistic adults on these platforms \cite{mazurek2013social}. Prior CSCW work shows how these technologies enable autistic individuals to connect with others and empower them to act more authentically \cite{van2023understanding}. At the same time, autistic social media users have found it difficult, and expressed a desire, to understand the social norms around using social media \cite{van2023understanding}. Also challenging is managing sensory sensitivities, cognitive load, and anxiety coming from using social media \cite{zolyomi2019managing}. Social media is often designed to privilege neurotypical values. For example, many mainstream platforms are built for phatic interactions, or social pleasantries, while autistic individuals have been found to favor interest-based sociality \cite{barros2023my}. In some cases, technologies are explicitly designed to train normative social behaviors in place of neurodiverse ones \cite{williams2020perseverations}. Researchers have worked with members of this population to envision improved social media features \cite{barros2023my} and have noted certain technology affordances that can be beneficial (e.g., captioning, meeting recordings, less time pressure for responding) \cite{das2021towards}.

As a response to this neuro-normative sociotechnical trend, researchers have developed specialized social networking platforms for autistic users. Hong et al.'s \textit{Social Mirror} \cite{hong2012designing} platform connects autistic individuals with a trusted set of family, friends, and professionals for the purpose of asking questions and getting advice. The platform catered to their preferred interaction style by allowing users to initiate conversations through pictures and visual representations. However, safety and privacy was a key concern participants noted about utilizing their social network. Additionally, research has also found that users can feel "othered" when using platforms that are specifically designed for neurodiverse users \cite{zolyomi2018values}. Autistic communities have also evolved on more general platforms such as Minecraft. Such online communities support interest-driven interactions mediated through various channels (e.g., chats, videos, forums) resulting in meaningful social interactions \cite{ringland2016will}. This work highlights how certain styles of mediated interactions can be beneficial for autistic individuals when optimized for their use. Given the benefits of using mainstream social media platforms and connecting with \finaledit{others} in the general population, it is important to help autistic individuals navigate these platforms safely.}
 
\subsection{Social Media \edit{and Privacy Harms for} Autistic Users}
 \edit{Social media privacy research with the general population has shown that users often orient their behavior around injunctive and descriptive norms \cite{morgan2018welcome}, and that violating these norms can lead to social backlash \cite{rost2016digital}. \finaledit{Moreover}, many sociotechnical systems are built around narrow cognitive and communicative norms, which can produce heightened risks and harms for users whose interaction styles do not align with these expectations \cite{page2022perceiving}.} Given the affordances introduced by social media \edit{(visibility\finaledit{,} editability, persistence, and association} \cite{treem2013social, bucher2018affordances}), research has uncovered many issues related to self-disclosure and self-presentation \cite{xu2008examining, page2022social}. Users often experience context collapse where family, friends, professional connections, and acquaintances all have equal access to them and their information \cite{marwick2011tweet}, or they overestimate their privacy by underestimating their audience size \cite{litt2012knock, litt2016imagined, binder2009problem, marwick2011tweet}. Social media users try to protect their privacy, such as withholding and selectively disclosing information \cite{consolvo2005location, wiese2011you, xu2009effects} or using privacy settings and rejecting or deleting connections \cite{acquisti2006imagined, ellison2011negotiating, debatin2009facebook, vitak2012won}. They may even create a separate social media account for each audience \cite{dennen2017context, page2022social}. 
 
 These \finaledit{challenges} highlight the need for increased digital and privacy literacy for \edit{all} social media users\edit{, but in particular for vulnerable users, who are disproportionately subjected to harm}. \edit{In addition to facing such privacy challenges, autistic individuals are also} more likely to face negative outcomes on social media platforms, such as relationship damage, physical harms, and cyberbullying \cite{page2022perceiving, triantafyllopoulou2022social}. Page et al. found that this can be attributed to differences in the perceived affordances of social media \cite{page2022perceiving}. For instance, autistic adults often imagined \finaledit{very} small audiences and thus information meant for a specific person was often shared widely across their whole network. Recent work suggests that autistic social media users do sometimes employ privacy practices such as self-monitoring, but not consistently and without the nuance to adapt to different situations \cite{cullen2024towards}. \oldedit{In the tradition of CSCW researchers who have examined social media privacy practices of vulnerable populations such as older adults \cite{murthy2021individually}, LQBTQ+ individuals, \cite{devito2018too}, and youth \cite{wisniewski2015preventative}, \edit{we also need to study the privacy practices of autistic individuals}.} 

\subsection{\edit{Digital Literacy and Privacy Education in HCI and Autism-Specific Educational Approaches}} 
Digital privacy education has been widely studied in HCI, primarily for the general population and adolescents \cite{stein2020learning, smith2023privacy, khan2024teaching, kumar2020strengthening}. Topics include online disclosure, privacy settings, and AI-related harms such as tracking and personalization \cite{egelman2016teaching, khan2024teaching, stein2020learning, smith2024know}. CSCW researchers have found privacy education games effective for children and adults to assess and articulate privacy norms \cite{blinder2024evaluating, berkholz2025playing}. Classroom-based K-12 and post-secondary privacy education has increased student awareness of and desire to manage privacy \cite{egelman2016teaching, hardin2020digital, khan2024teaching}. There has been some research into library-based digital citizenship courses for autistic individuals, but no digital privacy education \cite{phillips2020cyberbullying}.

Since The Salamanca Statement released by UNESCO advocating for equal access to education, there has been an increased emphasis to include neurodiverse and disabled populations in mainstream education, which can be mutually beneficial \cite{white2023creating, ferraioli2011effective}. However, autistic students face challenges with socialization, communication, and academic learning in mainstream education settings \cite{white2023creating, ferraioli2011effective, roberts2016review, bellini2010strength, burnham2017remediating, saggers2016australian}. Neurodiverse students have been found to have lower \finaledit{academic} performance \cite{mcleod2019experiences}. Heavy reading loads, lengthy written assignments, and collaborative group work pose greater challenges to autistic learners \cite{mcleod2019experiences, roberts2010topic, fleury2014addressing, mason2008theory, nation2006patterns}. While work is emerging on autistic learners in university contexts \cite{sarrett2018autism, cai2016educational, mcleod2019experiences}, the majority of education research focuses on children K-12 \cite{iovannone2003effective, chaidi2020parents, fleury2014addressing, lynch2009inclusive}. \edit{We study transition-aged young adults (18–30) not attending college, an understudied group with substantially lower economic, social, and physical health indicators than the general population \cite{warfield2014transition, maslahati2022adults, croen2015health}.} 

CSCW research has identified best practices for autistic learners. Content must be developmentally appropriate and align with student interest \cite{iovannone2003effective, sarrett2018autism, lowy2023building}, presented through a structured educational routine with systematic instruction, yet be somewhat flexible and individualized \cite{iovannone2003effective, sarrett2018autism}. Mentoring from educators and educational systems should come with a deep understanding of autistic behaviors and perspectives \cite{cai2016educational, roberts2016review, ness2013supporting, adreon2007evaluating}. Importantly, incorporating rule-governed behaviors can steer autistic individuals towards  \edit{making choices that minimize harms}~\cite{bradley2022rule, tarbox2011rule}. Rule-governed behaviors are generalized rules that determine social behavior and can be applied to a variety of social circumstances, promoting independence \cite{bradley2022rule}. \edit{Leveraging these best practices, our course incorporates rule-governed privacy protective behaviors, our content is aligned with relatable experiences, and we work within the bounds of autistic preferences (for social interactions, communication, etc.).}

\section{\edit{Privacy Rules for Inclusive Social Media (PRISM): An Educational Intervention for Autistic Users}}

\edit{We designed Privacy Rules for Inclusive Social Media (PRISM), a classroom-based privacy education program tailored to the learning styles and lived experiences of autistic young adults with substantial support needs. This section describes the design rationale and structure of PRISM, including the pedagogical principles and neuro-affirming framework that guided its development. We then detail the six instructional modules that operationalize these principles through scenario-driven, platform-specific privacy education.}

\subsection{Design Considerations and Overview of the PRISM Education Program}
\edit{PRISM is a classroom-based social media privacy education program designed as a sequence of structured, scenario-driven learning modules for autistic young adults. As a baseline, we drew upon existing social media privacy education research in HCI that has examined how individuals gain privacy awareness through classroom instruction, interactive activities, and scenario-based learning \cite{egelman2016teaching, hardin2020digital, blinder2024evaluating, berkholz2025playing}. The curriculum was then developed over eight months through iterative collaboration with a multidisciplinary team of autism researchers, autistic young adults, and community advocates, and was designed to be accessible across a wide range of cognitive and literacy profiles \cite{iovannone2003effective, sarrett2018autism, adreon2007evaluating}. PRISM is organized into six instructional modules delivered through weekly class sessions. Each module targets a distinct set of privacy-related challenges on social media and follows a consistent instructional flow: participants first complete a brief pre-assessment, engage with structured instructional content interspersed with group discussion and activities, and then complete a post-assessment using the same format. We translated evidence-based and neuro-affirming practices into the design of PRISM by adopting the following overarching design principles:}

\edit{
\begin{itemize}
    \item \textbf{Rule-based decision support.}  
    PRISM uses explicit, scenario-driven and rule-governed frameworks \cite{tarbox2011rule} to support privacy decision-making in complex social media contexts (Figure~\ref{fig:rulebasedslide}).

    \item \textbf{Consistent instructional structure.}  
    All sessions followed a stable instructional routine \cite{iovannone2003effective}, including pre-assessment, instruction, and post-assessment, using consistent formats and instructors to support predictability and reduce cognitive load \cite{ness2013supporting}.

    \item \textbf{Platform-specific, experience-grounded content.}  
    Instructional materials were grounded in participants’ lived experiences on commonly used platforms such as Facebook and Instagram \cite{page2022perceiving}, supporting inductive learning through concrete, familiar scenarios.

    \item \textbf{Accessible instructional design.}  
    Materials employed simplified language, visual scaffolding, and text written at approximately a fourth-grade reading level to accommodate diverse literacy profiles \cite{cerga2019improving,nally2018analysis,nation2006patterns,brown2013meta,ricketts2013reading}.

    \item \textbf{Support for nuanced privacy reasoning.}  
    The curriculum was designed to help participants move beyond rigid, all-or-nothing privacy rules toward more context-sensitive and flexible strategies \cite{varanda2017cognitive}.

    \item \textbf{Extensible, scenario-based format.} 
    PRISM adopts a structured, scenario-driven format \cite{gray2021social} that can be extended to other digital literacy topics such as misinformation, online identity, and safe content creation.
    A detail-focused cognitive style is common for autistic individuals \cite{happe2006weak}. Rather than teach participants high level privacy rules that they have to apply to different situations, we took an inductive approach. We created a variety of exemplar scenarios to represent nuanced variations of possible privacy situations (e.g., friend request with different pictures, names, or number of mutual friends). This allowed them to intuit privacy rules.

    \item \textbf{Co-design with autistic stakeholders.}  
    Autistic students, advocates, and a multidisciplinary team of autism researchers were involved throughout the design process to ensure developmental appropriateness and alignment with participant needs \cite{spiel2020nothing}.
\end{itemize}
}

\subsection{\edit{Grounded on Evidence-based Neuro-Affirming Educational Frameworks}}

\edit{Autistic individuals often perceive and interact with social environments differently than neurotypical individuals \cite{mottron2006enhanced}. For instance, autistic individuals often demonstrate strengths in systematizing and rule-based reasoning, supporting pattern recognition and explicit decision-making \cite{mottron2006enhanced, baron2017autism, brosnan2016reasoning}. As such, the goal of PRISM is not to reinforce the neuro-normative design of social media platforms, but rather to work with participants’ natural communication styles and reasoning strengths to support autonomy and self-determination in online decision-making. Two complementary instructional strategies (i.e., scenario-based learning through Social Stories and explicit rule-based decision support) operationalize this neuro-affirming approach.}

\subsubsection{\edit{Scenario-Based Learning through Social Stories}}

\edit{PRISM draws inspiration from Carol Gray’s Social Stories \cite{gray2021social}, which are short narratives designed to describe social situations and expectations in an accessible and non-prescriptive manner. Rather than instructing individuals on how they should behave, Social Stories provide contextual information that enables learners to anticipate situations, understand possible responses, and consider different outcomes. When used to enforce social conformity, Social Stories can be harmful; however, when carefully written to explain expectations without mandating compliance, they can support autonomy and self-determination \cite{camilleri2025self}. 
In PRISM, scenario-based materials function as contextualized Social Stories that describe common social media situations (e.g., receiving a friend request, being tagged in content, or asked for personal information). These scenarios help participants understand platform norms and affordances, reducing uncertainty and anxiety associated with implicit expectations. Importantly, PRISM explicitly affirms participants’ agency to choose whether to follow or disregard these norms, recognizing that understanding social expectations can be empowering even when individuals opt not to conform \cite{van2023understanding}.}

\subsubsection{\edit{Rule-Based Decision-Making as Cognitive Support}}

\edit{In parallel with scenario-based learning, PRISM leverages autistic learners’ strengths in rule-based and systematizing reasoning \cite{mottron2006enhanced, baron2017autism}. Rather than relying on implicit social inference or trial-and-error learning, instructional content provides explicit decision criteria that participants can apply when navigating privacy-sensitive situations. These rule-based frameworks support consistent reasoning across contexts, particularly in complex or ambiguous social media interactions. By combining rule-based decision support with scenario-driven contextualization, PRISM enables participants to reason about privacy choices in ways that align with their cognitive preferences while preserving flexibility and agency. This approach supports informed decision-making without prescribing a single “correct” way to engage with social media, reinforcing PRISM’s broader goal of empowering autistic users to navigate online platforms on their own terms.}

\subsection{\edit{Detailed Design of the PRISM} Educational Modules}
\edit{The PRISM program} was designed as 6 ordered modules: (1) \edit{Platform Rules and Typical Use}, (2) Choosing Safer Privacy Settings, (3) How to Identify Fake Profiles, (4) Types of Social Groups on Social Media, (5) \edit{Safe} Interactions Based on Social Group, and (6) Social Media vs Reality. Modules 1, 3, 4, and 5 follow a rule-based approach. \edit{These modules walked students through the concepts and included slides with discussion questions which students could answer if they chose to.} The two modules, Choosing Safer Privacy Settings and Social Media Versus Reality are not rule-based as they are not teaching students about what privacy-sensitive decisions to make on social media. Instead, the Settings module is primarily focused on assisting students manually updating their social media settings, and the latter module teaches students about how other people use social media. \edit{These modules specifically taught about Facebook and Instagram, platforms whose privacy risks have been identified in prior research \cite{page2022perceiving}. Our field site staff confirmed that these are commonly used and led to privacy risks for our participants.} We now outline the reasoning behind teaching each of these modules.

\begin{figure}[htp!]
    \centering
    \includegraphics[width=1\linewidth]{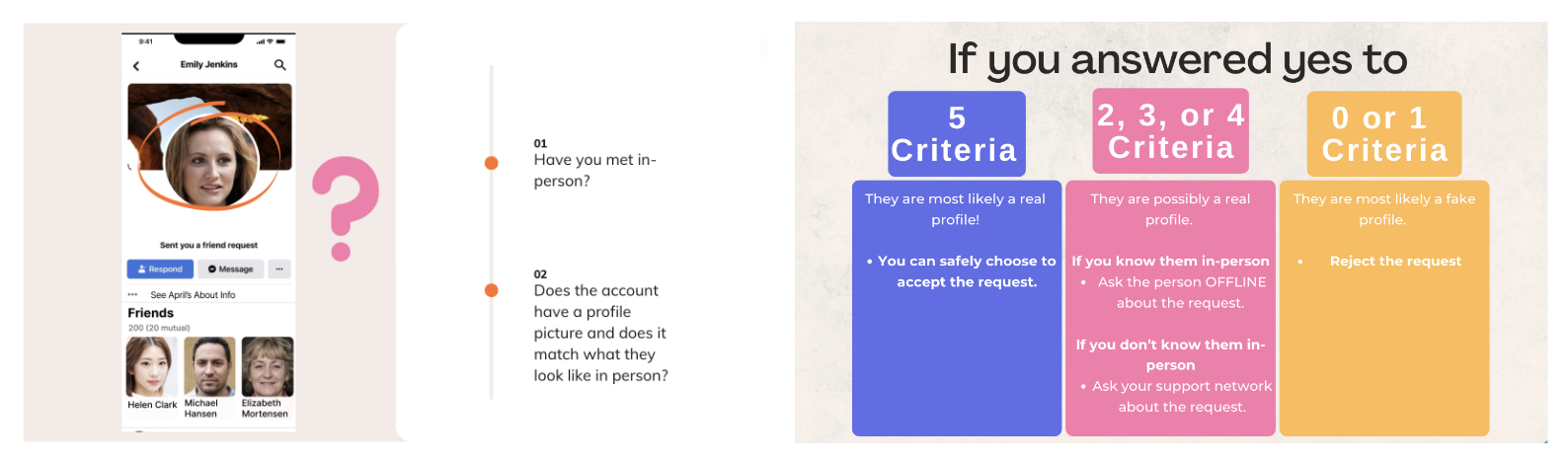}
    \caption{Example of Rule-Based Educational Slide. There were 5 rules that students were taught. 2 of the criteria are shown on the left screen. Based on the number of criteria which a profile met, we recommend whether the participant to accept, reject, or further examine the request (shown on the right screen)}
    \label{fig:rulebasedslide}
\end{figure}

\begin{figure}[htp!]
    \centering
    \includegraphics[width=1\linewidth]{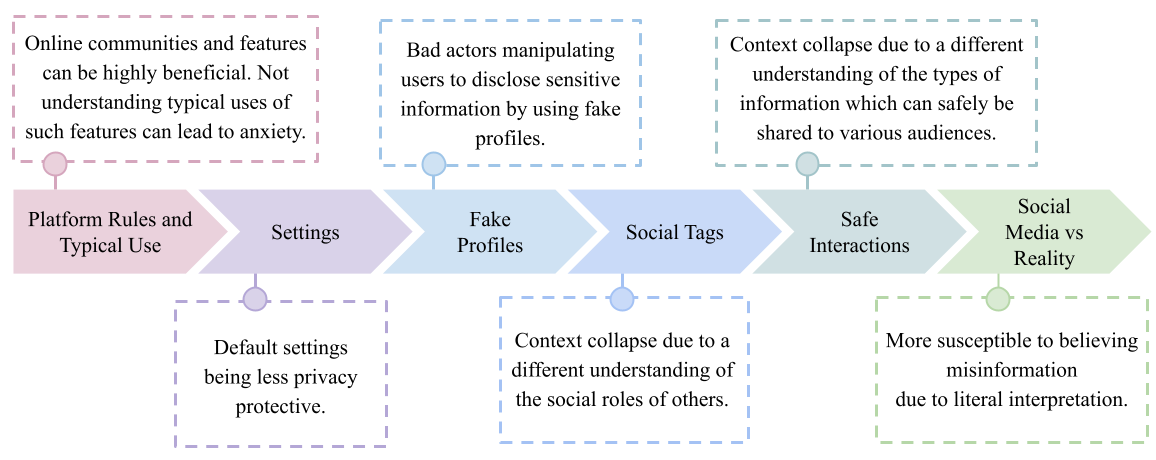}
    \caption{The Modules and the Challenges they Each Address}
\label{fig:rulebasedslide}
\end{figure}

\subsubsection{\edit{Platform Rules and Typical Use}}
\edit{Prior work has found that community building on social platforms can be highly beneficial for autistic individuals \cite{wang2020benefits}. One such reason for this is it allows them to connect with others and discuss the topics they are most interested in \cite{ringland2016will}. At our field site, the autistic young adults requested to learn about social media features and how people \finaledit{typically} used the platforms and said features. As such, we developed a set of educational materials which describe the explicitly-stated platform rules (e.g., Facebook only allows one profile per user), various features (i.e., page, group), accounts types (e.g., personal profile, 'Finsta', topic, verified, and business accounts), and typical ways people use these accounts. The communication features were based \finaledit{on} what was available on Facebook. Account types were  \finaledit{drawn from} a mix of what is native to Instagram (personal, verified, business accounts), account types examined in prior research ('Finsta' \cite{huang2022finsta}), and recommendations from a multidisciplinary team of autism researchers (topic accounts). We taught students about various positive uses of social media (e.g., connect with friends, run a business), and the platform policies. We were not prescriptive in telling students how they \textit{should} act on social media.}

\subsubsection{Choosing Safer Privacy Settings} 
Default settings are often less privacy protective \cite{watson2015mapping} and many of our students never changed their default settings. We pointed out more privacy protective settings, which they mostly chose to follow. We walked students through and made recommendations on the settings in Instagram and Facebook, assisting students as needed to make changes. This included controlling who can view personally created content (i.e., posts, stories, comments, likes, interests), personal information (i.e., date of birth, contact information, place or school or work), who can share personally created content (i.e., stories\finaledit{,} posts), and who can associate them with other content (i.e., tagging). Other topics included blocking or restricting accounts or content, settings for social media messaging, how to prevent unwanted messages or hide messages, and disabling message pop up.

\subsubsection{How to Identify Fake Profiles}
Prior work uncovered that autistic individuals may be more likely to interpret the ``friend'' label literally and accept friend requests from unknown or unsafe users \cite{page2022perceiving}. This module discusses the dangers of fake profiles and how to avoid them using a rule-based approach \cite{plummer2017introduction}. It provides students with an algorithmic process to decide whether or not to be ``Friends'' with someone on Facebook. Namely, students evaluate 5 criteria: (1) ``Have you met this person in-person?'', (2) ``Does the account have a profile picture and does it match what they look like in person?'', (3) ``Is this your first connection request from this person?'', (4) ``Do they have more than 50 friends and less than 2000 friends?'', (5) ``Do you have 5 or more mutual friends with this person?''. If a profile matches 5 of the criteria, then we teach students that they are most likely safe to be connected with. If a profile matches 2 to 4 criteria and they know the person offline, then we recommended that they ask the person about the request in-person or discuss it with someone in their support network if they don't know them offline. These criteria were developed in consultation with empirical studies characterizing fake profiles \cite{roy2021fake}.

\subsubsection{Types of Social Groups on Social Media}
 The ``Circles Curriculum'' is commonly taught to autistic individuals to describe social roles of individuals in an offline context \cite{circlescurriculum}. However, it does not take into account roles such as ``online friends'' nor account for someone taking on multiple roles. To help students more accurately classify people into roles on social media, we introduced ``social tags'' and demonstrated how to apply any number of tags to a given person's account on Facebook (i.e., close family, distant family, in-person friends, school peers, personal service providers, work peers, bosses and teachers, online friends, acquaintances, community helpers, and strangers).

\subsubsection{Safe Interactions Based on Social Group}
To mitigate context collapse harms \cite{page2022social}, we teach students the types of information which \edit{are safe} to share online with the previously taught social tags. Focusing on privacy sensitive or personally identifiable information \cite{krishnamurthy2009leakage}, information was grouped into more general (e.g., first name, age range, gender) and more specific (e.g., contact information, medical history, daily routine). We taught what information group students can \edit{safely} choose to share and what is \edit{safe} for certain people to ask for. Some information is never \edit{safe} to share on social media, which can be compromised (e.g., government-issued numbers, financial information, passwords, and home-layouts). \edit{Because autistic social media users often interpret requests as instructions to act \cite{page2022perceiving}, we reinforced that someone asking for information is not an imperative to do so - students can choose whether to share.}

\subsubsection{Social Media vs Reality}
Prior work has found that autistic social media users may be more likely to face emotional harms due to social media use \cite{page2022perceiving}. Reasons for this include fear of missing out (FOMO) and comparison to others. We taught students about how many people use social media only when they are happy and use photo editing. Furthermore, as prior work has found that autistic individuals may be likely to interpret content on social media literally \cite{page2022perceiving}, we also taught students about what misinformation is and how to avoid it. Strategies include fact-checking information found on social media, being cautious, being aware of capabilities like photo editing, and asking themselves why they think the information is true and why they would want to share it.

\section{Methods}We worked with a local residential program supporting transition-age autistic YAs (18-30's) in the Western United States. They offer classes to their residents and the community including a variety of life skills (e.g., cooking, budgeting) and interest-based (e.g., art studio, music) classes. Through their program, we offered a weekly in-person course that students could sign up to attend. \edit{An in-person course presents a chance for social interaction to support learning, minimized technical risks of online classes, and was accessible to students already enrolled in other classes at the field site.} We ran a weekly 50 minute class over a 14 week period, enrolling a total of 29 students. \finaledit{We taught one of the six privacy modules every other week, with breaks aligning with the field site's academic calendar.} \edit{We limited class size to a maximum of eight students to alleviate sensory overload and social anxiety (which commonly co-occurs with autism) \cite{white2009anxiety, bellini2006development}}. All students were YAs (ages 18 - 30's) with level 2 autism support needs \cite{weitlauf2014brief}. \finaledit{The DSM-5 defines level 2 as "requiring substantial support needs" \cite{dsm5}. Specifically, all of our participants used verbal communication, read at a fourth grade level, had access to internet, and lived in a residential community that offers support for daily living.} We chose to recruit this age group of autistic individuals who are most likely to be literate and online as a result of turn-of-the-century legislation that gave autistic individuals access to public school systems \cite{ADANationalNetwork2024}. All classes were taught by a team of multiple (3+) researchers \finaledit{with a member of the field site staff present}. We recorded audio transcripts of each class. Each student was compensated with a \$10 Amazon gift card for each class they attended, an amount recommended by our field site. \oldedit{\edit{We took an ethnographically informed approach, reading their program materials, talking with various staff members to understand the students and their prior experiences with social media, \finaledit{and} being involved in other educational events hosted by the field site.}} We detail our study design in the following sections.

\subsection{Design-Based Research Informed Methodology}
We use a design-based research approach \cite{reimann2010design, sandoval2004design} which has been used in HCI research for classroom-based technology interventions \cite{khan2024teaching}. Design-based research suggests that iterative design and research that takes place in-situ can lead to authentic knowledge \cite{armstrong2020design}. Design-based research is composed of 6 elements: focusing on who the learner will be, understanding the learner and the existing solutions, defining goals, conceiving the solution, building the solution, and testing the solution \cite{easterday2014design}. \finaledit{We} met with the field site staff members to understand the needs of students. For instance, staff told us the importance of having a module discussing interpersonal comparison (our Social Media vs Reality module). Based on these preliminary discussions and \finaledit{prior research} \cite{page2022perceiving, cullen2024towards}, we were able to develop our classroom materials and then deploy them in a classroom setting. This methodological framework is ideal for our research as we are dealing with a unique context which differs greatly from a contained laboratory environment. This framework also allows for the necessary flexibility in understanding and working with this population.

\subsection{Pre and Post Intervention Assessment}
This study implemented a pre-assessment and post-assessment design. This approach has been utilized by many studies to evaluate educational interventions by capturing initial understanding of a topic through an assessment, providing an intervention, then following up with a similar or the same assessment to capture final understanding of the topic. The pre and post tests are compared to analyze the success of the intervention \cite{choi2022influence, arifa2023pbl, pozzan2024experimental, herlambang2023project, choi2024influence, haveyourcake, Crystallize, CanChildrenUnderstandML, ChildrenADHD, khan2024teaching}. Conducting assessments immediately before and after the intervention is a valid approach \cite{haveyourcake, Crystallize}. Prior work has found that autistic individuals have greater difficulty with short-term memory \cite{desaunay2020memory, southwick2011memory, cheung2010verbal} and so we sought to examine whether this population would remember content in the short-term. We chose to not include a control group for several reasons. First, our focus is on within-subject change and how much each individual student improves \cite{alam2019comparative}. Second, prior literature \cite{page2022perceiving} shows that this population is at increased harm due to social media use, and we felt it would be unethical to withhold a potentially beneficial intervention. The best practices for ethical HCI research state that when working with vulnerable populations the interventions must provide a greater benefit to participants than to the researchers \cite{brown2016five}. Finally, as we are working with a residential facility where students pay to attend classes, it would be unethical to run a 14-week control group where students do not learn online safety.

In each of our sessions, participants were given a knowledge assessment immediately before and after the educational intervention. Both assessments were identical and students were not told whether they had correctly or incorrectly answered any questions. \oldedit{Additionally, participants were encouraged to answer the same way if they felt their answer was correct the first time.} Questions on the assessment were multiple-choice, including true/false responses, single answer responses, and multiple answer responses. The assessment contains 5 to 12 questions, depending on the session. These assessments were, on average, completed under 5 minutes. The specific questions asked on the assessment are included in appendix ~\ref{app:prepost}. Generalizing concepts has been found to be a challenge for autistic individuals \cite{brown2012generalization} since slightly different scenarios may be viewed as completely different. As such, many of the questions on the assessment are only slightly different. \edit{The questions include scenarios (e.g., "A [residential facility] peer asks for your personal email") that span the range of possible rule combinations, grounded in social situations relevant to our participants' lives. The situations were determined based on input from field site staff and a multidisciplinary autism research group.} \finaledit{They allowed} us to test participants' understandings of specific nuances. 

\begin{figure}[htp!]
    \centering
\includegraphics[width=0.75\linewidth]{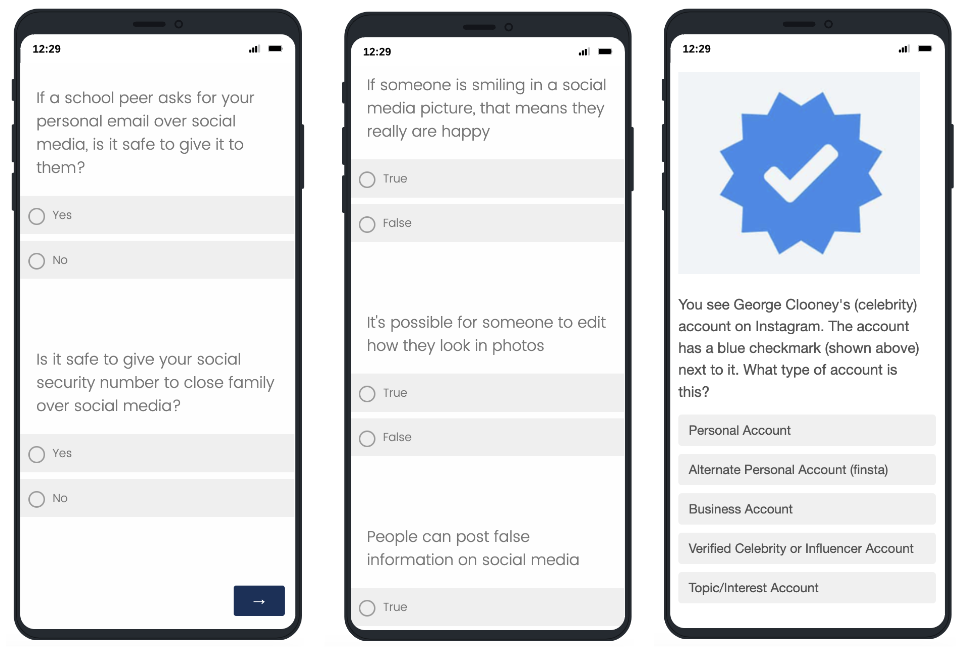}
    \caption{An example of the pre and post assessment questions}
\label{fig:prepostexample}
\end{figure}

\subsection{Ethical Considerations}
In this study, we were examining the potentially sensitive topic of social media use and the harms which arise from it for autistic YAs. As such, we considered the ethical implications of our study and ways to mitigate potential harms of participation. In the development of our educational materials, we collaborated with autism researchers and autistic individuals to ensure that our materials would not be harmful for participants. All materials were also created to promote the most privacy preserving behaviors. A teacher from the program attended each class. If students ever wanted to leave class early or not participate in the pre-post assessments they were free to do so. \edit{In order to reduce feelings of distress due to forced socialization, we did not pressure students to verbally contribute to classroom discussions.} All transcripts were anonymized and student performance data (pre and post assessment scores) were not shared with the field site. All students were their own guardians, and we obtained written informed consent from all of the students. This study received IRB approval from the first author's institution.  

\subsection{Positionality Statement}
Most members of the research team have family members on the autism spectrum and/or are experts working with autistic individuals, including one author who self identifies as autistic. We drew from the team's lived experiences to inform the design of the educational intervention and better understand student interactions and insights. Namely, these lived experiences aided the authors in understanding and supporting the communication style of the participants in the design of the materials and analysis of data. For instance, researchers were aware of not including extraneous content or \finaledit{distracting} designs on the slides and the need for literal language. In the development of materials, researchers were also mindful of structuring lessons into manageable \finaledit{short} segments.

\subsection{Data Analysis Approach}
\subsubsection{Analyzing Student Utterances (RQ\oldedit{1})} We analyzed the audio transcription of what was discussed in class. We engaged in an iterative inductive coding process \cite{strauss1987qualitative} to generate themes shared by students about privacy in the context of a classroom intervention. The first author initially coded all utterances to identify emerging codes about privacy management and create the codebook. In the second round of analysis, the first and third author then collaboratively coded all utterances. The second author was consulted to reconcile any differences between the coders. \oldedit{Coders reconciled any differences in coding through discussion until reaching consensus, a best practice in CSCW qualitative research \cite{mcdonald2019reliability}. They determined they had reached coding saturation when coding the final semester of classes, and no new codes emerged}. In table ~\ref{tab:rq2qual} we present our final codebook which identified three overarching themes that students discussed during class.

\begin{table}[] \edit{
\begin{tabular}{|>{\raggedright\arraybackslash}p{0.07\linewidth}|>{\raggedright\arraybackslash}p{0.3\linewidth}|>{\raggedright\arraybackslash}p{0.55\linewidth}|}
\hline
\textbf{Theme} & \textbf{Code}                                                      & \textbf{Quote} \\ \hline
\multirow{3}{*}{\begin{tabular}[c]{@{}l@{}}\textcolor{black}{Strong} \\ \textcolor{black}{Privacy} \\ \textcolor{black}{Beliefs}\end{tabular}} &
  Anticipation of Online Rejection &
  \textcolor{black}{\textit{``[People create fake profiles to] Scam, cyber bully, be a predator, steal people’s information.'' (P6, \oldedit{Fake Profiles})}} \\ \cline{2-3} 
               & A Sense of Vulnerability Resulting from Concerns of Potential Harm& \textcolor{black}{\textit{``There are some things you just shouldn’t tell your boss.'' (P18, \oldedit{Social Tags})}}        \\ \cline{2-3} 
               & A Sense of Invulnerability Resulting from Not Foreseeing Risk& \textcolor{black}{\textit{``A lot of times I’ve like, posted like, just, like some rant about something, like, just kind of venting... And like, and then I don’t really care. Like, I’m like, Okay, anyone I know can see it. I don’t really care, right? Like, I’m not, like, trying to hide, like, details, like, I’m just very open. I guess maybe that’s a bad thing'' (P17, \oldedit{Social Tags})}}        \\ \hline
\multirow{4}{*}{\begin{tabular}[c]{@{}l@{}}\textcolor{black}{All} \\ \textcolor{black}{-or-} \\ \textcolor{black}{Nothing} \\ \textcolor{black}{Privacy} \\ \textcolor{black}{Actions}\end{tabular}} &
  Reducing Cognitive Load from Social Interaction through Privacy Features &
  \textcolor{black}{ \textit{``Block anyone you don’t know '' (P12, \oldedit{\edit{Platform Rules and Typical Use}})}} \\ \cline{2-3} 
               & Only Connected with Specific Established Offline Relationships     & \textcolor{black}{\textit{``That’s why my accounts only for family and friends'' (P2, \oldedit{Fake Profiles})}}        \\ \cline{2-3} 
               & Social Avoidance through Platform Management                       & \textcolor{black}{\textit{``[I’m on Instagram,] because all my weird family members are on Facebook and they’re not on Instagram. So then they don’t see {[}what I post{]}'' (P14, \oldedit{\edit{Platform Rules and Typical Use}})}}        \\ \cline{2-3} 
               & Considering Certain Types of Information as Always Unsafe to Share & \textcolor{black}{\textit{``That’s why just don’t send any personal information on the internet, even if it’s with a family member'' (P24, \oldedit{\edit{Safe} Interactions})}}       \\ \hline
\end{tabular}}
\caption{Codebook for Student's Classroom Utterances}
\label{tab:rq2qual}
\end{table}

\subsubsection{Learning Outcomes (RQ\oldedit{2})}
To test the effectiveness of the six educational modules we calculated a pre-assessment score and post-assessment score for each module. \finaledit{These} scores were calculated by dividing the number of questions answered correctly by the number of questions in each module, giving a standardized score between 0 and 1. We hypothesized that each of the designed learning modules would lead to an increase in participants\finaledit{'} assessment scores. Using the Shapiro-Wilk test, we first checked the normality of the data. The results of these tests are shown in Table {\ref{tab:quantresults}. We find that three of the module assessments (Fake Profiles, Social Tags, Social Media vs. Reality) were normally distributed and the remaining (\edit{Platform Rules and Typical Use}, Settings, Social Tags) were not. Following these tests, we then used either a two-sided paired t-test (when the data was normally distributed) or the Wilcoxon-Signed Rank Test \cite{woolson2005wilcoxon} to assess the impact of the educational module on the assessment scores. All of the statistical analyses were completed with the STATA 17.0 statistical software. \oldedit{To better understand these quantitative results, we revisited the data to find any qualitative trends that helped us explain them and present those in our findings.}

\subsubsection{\oldedit{Student Outcome Analysis (RQ3)}}
To examine how student behaviors aligned with performance, we examine classroom behaviors during the Fake Profiles\edit{, Social Tags, and Safe Interactions modules. We chose to examine these modules as they were all statistically significant and followed a similar classroom format. We did not choose to examine the \edit{Platform Rules and Typical Use} module because, although statistically significant, it occurred during the first class session which was primarily focused on rapport-building and introductory discussion, making it less comparable to subsequent modules in terms of classroom dynamics. For each class,} we divide students into three groups: 1) Students who had a positive change in assessment score; 2) Students who had no change in assessment score; and 3) Students who had a negative change in assessment score. The first author initially open-coded these student's classroom utterances to better understand their behavior. The third author then independently coded these responses. Differences in codes were reconciled through discussion, a qualitative research best practice \cite{mcdonald2019reliability}. The first author then examined how these codes varied between group. For some of the classes, we did not record the lesson due to student consent. We will be reporting percentages from the students who were audio recorded.

\section{Results} 
\oldedit{Participants discussed many of their privacy beliefs and actions in the classroom setting\finaledit{, which enabled us to answer RQ1. We answer R2 by using our pre- and post- assessment scores to show that the} intervention was effective in teaching participants about online social media privacy. \finaledit{Finally, we answer RQ3 by identifying} specific classroom behaviors that were indicative of student success. \finaledit{Here we present our results.}}

\subsection{How Autistic YAs Discuss Privacy \oldedit{and Pre-existing Behaviors} in a Classroom Context (RQ1)}

\subsubsection{Privacy Beliefs}
\edit{There is a diversity of strong privacy beliefs among participants. On one hand, participants had major concerns regarding their privacy and online safety stemming from beliefs that others would harm them. \finaledit{On the contrary}, other participants seemed to feel a sense of invulnerability to privacy harms.} This will be further discussed in the following section.  

\paragraph{\edit{Anticipation of Online Rejection}}
\edit{Participants anticipated that they were going to be rejected or harmed by others online. Namely, participants would share concerns of bad actors having malintent and harming them physically, emotionally, or financially. }Many students noted that they did not trust anyone that they haven't met in-person. A few students even shared that they followed the rule to \textit{``block anyone you don't know''} (P12, \oldedit{\edit{Platform Rules and Typical Use}}). Students also stated that they had no desire to talk to anyone that they don't know personally. Often times, students felt like there was no reason to interact with that person, \textit{``if you don't know that person, then there's no point in answering it''} (P2, \oldedit{Fake Profiles}). In other cases, students just had a rule for themselves not to talk to the stranger. For instance one student shared, \textit{``I don't talk to anyone online that I haven't met personally''} (P1, \oldedit{Social Tags}) while another student similarly shared, \textit{``I'm running into this all the time where I'm having to not follow or not reply to profiles because I know that they could possibly be fake.''} (P10, \oldedit{Fake Profiles}). 

Many participants also noted that they were concerned about others taking advantage of them. Some participants specifically noted that they were concerned about physical harms due to other social media users. For instance, one participant shared \textit{``Yeah, someone could come to your home address and stalk and like hurt you in some way''} (P21, \oldedit{\edit{Safe} Interactions}). While another participant similarly shared concerns regarding others knowing their physical location, \textit{``If you were going to show where you go to school, you're basically doxxing yourself''} (P23, \oldedit{\edit{Safe} Interactions}). Other students did not note physical concerns, but did mention concerns regarding financial scammers. For instance, one student shared, \textit{``They make you... they tell you all these things that umm... they're trying to get money out of you... a sweetheart scammer. My aunt has fallen victim to that before.''} (P9, \oldedit{Fake Profiles}). Another participant similarly shared, that \textit{``if [bad actors] take their money it ruins them and they don't have anywhere to live.''} (P1, \oldedit{Fake Profiles}). Some participants also noted concerns regarding other users being mean or cyberbullying them online. For instance, one participant shared that they would want to limit who can see their posts\finaledit{:} \textit{``So you protect yourself from people who want to be mean''} (P9, \oldedit{Settings}). Another student similarly shared that they would use the taught settings to, \textit{``Avoid the mean people and trolls''} (P10, \oldedit{Settings}).

\paragraph{\edit{A Sense of Vulnerability Resulting from Concerns of Potential Harm}}
Some participants specifically noted that there are things which you should not tell certain individuals based on the relationship you have with them. For instance, one participant noted that you shouldn't disclose \finaledit{interests} to classmates that you don't know well: \textit{``[If you share] a bunch of interests with like, some random classmates they're gonna be like, a little confused because they don't know you''} (P16, \oldedit{Social Tags}). Students more frequently mentioned being wary of disclosing information specifically to \finaledit{employers}. For instance, one participant shared, \textit{``There are some things you just shouldn't tell your boss''} (P17, \oldedit{Social Tags}). Other participants mentioned \finaledit{taking} precautions \finaledit{and concerns about} sharing information with \finaledit{authorities in the workplace}. \edit{These students noted that they felt like bosses would automatically fire them for certain posts\finaledit{:}} \textit{``Some bosses are stupid and they'll just fire someone because of one small thing''} (P16, \oldedit{Social Tags}). One participant shared their rule-based advice regarding managing connections with \finaledit{supervisors:} \textit{``Just block your boss on social media''} (P21, \oldedit{Social Tags}).

\paragraph{\edit{A Sense of Invulnerability Resulting from Not Foreseeing Risk}}
Some participants noted that \finaledit{they} \edit{felt a sense of invulnerability to privacy harms.} As our participants resided in a residential facility, some mentioned that they didn't feel as concerned about online others knowing their address. \finaledit{One student shared why }they wouldn't be as concerned\finaledit{:} \textit{``If I lived in a random house, then it would be different. But because I live with an apartment of people'' [I'm not as concerned]} (P17, \oldedit{Social Tags}). Other students mentioned that they aren't worried because they trust other social media users or don't feel like they ever share overly sensitive information\finaledit{:} \textit{``I trust everyone that's following me on Instagram and Facebook... Even if I don't know them that well, like, the information I'm sharing isn't like, private enough to really matter''} (P16, \oldedit{Social Tags}). Finally, other participants noted that they are very open and had the tendency to vent online. \textit{``A lot of times I've like, posted like, just, like some rant about something, like, just kind of venting... And like, and then I don't really care. Like, I'm like, Okay, anyone I know can see it. I don't really care, right? Like, I'm not, like, trying to hide, like, details, like, I'm just very open. I guess maybe that's a bad thing''} (P16, \oldedit{Social Tags})\oldedit{.}

\subsubsection{All-Or-Nothing Privacy Actions}
Overall, we found that participants often employed dichotomous behaviors when it came to privacy management. In the following sections, we discuss how this \finaledit{manifested} among participants.

\paragraph{\edit{Reducing Cognitive Load from Social Interaction through Privacy Features}}
Many participants noted that they would utilize blocking and other privacy features in their social media use. Some students noted that they would use these features after they had already incurred privacy violations. For instance, one participant similarly shared, \textit{``Somebody spammed me some weird stuff and I didn't want to deal with it. So screw you, blocked.''} (P1, \oldedit{Settings}). Other participants noted that they would proactively use privacy features to keep themselves safe online \edit{and reduce potential cognitive load}. Some students noted following dichotomous rules in how they would use these features. For instance, one student shared, \textit{``Block anyone you don't know.''} (P12, \oldedit{\edit{Platform Rules and Typical Use}}).

\paragraph{\edit{Only Connected with Specific Established Offline Relationships}}
While many of our students were mistrusting of all strangers, many of them also appeared to be fully trusting of their friends and family. For instance, when asked about changing settings to limit \finaledit{commenting} to only friends, one student replied that the reason it is safe to do so is \textit{``because your friends are your circle of trust''} (P1, \oldedit{Settings}). Similarly, many students noted that they are only connected with real-life friends and family on social media. For example, one student shared, \textit{``That's why my account's only for family and friends''} (P2, \oldedit{Fake Profiles}) while another student shared,\textit{ ``I actually make sure that all the people that I'm friends with I know in real life. Many of the people I'm friends with are family members''} (P9, \oldedit{Fake Profiles}).

\paragraph{\edit{Social Avoidance through Platform Management}}
Participants also noted that they would prevent privacy violations by managing the content they share and the platforms they use. Namely, participants shared that they would intentionally use one social media platform over another in order to avoid certain users. One participant shared that they only posted on Instagram \textit{``because all my weird family members are on Facebook and they're not on Instagram. So then they don't see it''} (P13, \oldedit{\edit{Platform Rules and Typical Use}}). Other participants similarly shared that they would choose one platform over the other based on the extent to which they were able to manage their audiences\finaledit{:} \textit{``I really like using Facebook because I prefer not having any people I don't know on any social media account... Facebook allows you to choose who you want to be friends with and on Instagram you can still choose, but like, these random people can just follow you.''} (P3, \oldedit{Fake Profiles})

\paragraph{\edit{Considering Certain Types of Information as Always Unsafe to Share}}
Some participants noted that they would control disclosure in various ways in order to protect or maintain their privacy. For instance, one \finaledit{participant} shared that they wouldn't post anything online in order to maintain privacy. Namely, when discussing privacy violations, the participant shared, \textit{``I don't really post anything, so I can't relate``} (P23, \oldedit{Social Tags}). Another participant shared that they would lie about information they deemed to be sensitive\finaledit{:} \textit{``Personally, I just lie about the year``} (P9, \oldedit{\edit{Safe} Interactions}). Other participants would disclose some information, but other pieces of information they deemed too sensitive to share with anyone. For instance, one participant shared, \textit{``That's why I just don't send any personal information on the internet, even if it's with a family member``} (P24, \oldedit{\edit{Safe} Interactions}). Another participant provided the reasoning for this as, \textit{``I still wouldn't share because then you could get hacked and they can figure that out``} (P21, \oldedit{\edit{Safe} Interactions}).

\paragraph{\edit{Other Privacy Violations}} 
Other students either didn't realize that there were actions they could take to avoid privacy violations or were unable to avoid privacy violations. For some students, the violation that occurred was a stranger trying to communicate with them\finaledit{:} \textit{``You don't want strangers commenting on your posts. I've dealt with that before. It's really weird.'' (P10, \oldedit{Settings})}. Other students faced potentially more severe privacy violations where bad actors actively pursued them or those around them. For instance, one student shared, \textit{``They tell you all these things... they try to get money out of you... a sweetheart scammer. My aunt has fallen victim to that before.''}(P9, \oldedit{Fake Profiles}) \finaledit{Another} student similarly shared, \textit{``Yeah. I had a stalker... She created like 15 different accounts to talk to me. She would put a different name on some of them.''} (P12, \oldedit{\edit{Platform Rules and Typical Use}}). These students did not explicitly mention how or whether they were able to mitigate these privacy violations. 

\subsubsection{\oldedit{Rigid Rules of Students Who Previously Faced Harms}}
\oldedit{Notably, informal conversations outside of class indicated that some participants recurrently faced privacy harms. For instance, one participant shared that they had previously connected with and messaged actors that they did not know, which led to restrictions on their technology access. During class, this student would often share rigid rules that they believed they followed\finaledit{:} \textit{``I actually make sure that all of the people that I'm friends with I know in real life. Many of the people I'm friends with are family members.'' (P9, Fake Profiles)}. Yet when showing their friends list to researchers, there were many signs that their connections were either fake profiles or bad actors. This was concerning as this participant had shared that they felt comfortable disclosing personal information to those they were messaging, \textit{``Like if you're really currently talking to them on your messenger, then [it is] probably the right time to share personal information.'' (P9, \edit{Safe} Interactions)}. Furthermore, this student shared that they had been messaging actors\finaledit{:}  \textit{``I've had to block someone because he wasn't the person I thought he was. He was trying to get me to send inappropriate pictures of my body. I told him no. And when he didn't back down, I had blocked him.''(P9, Settings)}. If this actor hadn't repetitively asked for explicit photographs, then this student may not have blocked them. This indicates that while the participant applied rigid rules that friends must be people they know offline, the rules did not help them understand what to do when someone fell outside of this context (e.g., the person is a famous actor rather than a friend).}

\subsection{Measured Learning Effects (RQ2)}

\begin{table}[htp!]
\begin{tabular}{|>{\raggedright\arraybackslash}p{0.1\linewidth}|l|l|ll|ll|ll|ll|>{\raggedright\arraybackslash}p{0.1\linewidth}|}
\hline
\multirow{2}{*}{Module} &
  \multirow{2}{*}{n} &
  \multirow{2}{*}{\begin{tabular}[c]{@{}l@{}}Total \# of\\ Questions\end{tabular}} &
  \multicolumn{2}{l|}{\begin{tabular}[c]{@{}l@{}}Normality\\ Test \textasciicircum{}\end{tabular}} &
  \multicolumn{2}{l|}{Mean \textasciicircum{}\textasciicircum{}} &
  \multicolumn{2}{l|}{Median \textasciicircum{}\textasciicircum{}} &
  \multicolumn{2}{l|}{St. Dev \textasciicircum{}\textasciicircum{}} &
  \multirow{2}{*}{Diff Test \textasciicircum{}\textasciicircum{}\textasciicircum{}} \\ \cline{4-11}
 &
   &
   &
  \multicolumn{1}{l|}{Pre} &
  Post &
  \multicolumn{1}{l|}{Pre} &
  Post &
  \multicolumn{1}{l|}{Pre} &
  Post &
  \multicolumn{1}{l|}{Pre} &
  Post &
   \\ \hline
\textbf{\edit{Platform Rules and Typical Use}} &
  \textbf{25} &
  \textbf{10} &
  \multicolumn{1}{l|}{\textbf{0.92*}} &
  \textbf{0.62***} &
  \multicolumn{1}{l|}{\textbf{0.77}} &
  \textbf{0.88} &
  \multicolumn{1}{l|}{\textbf{0.8}} &
  \textbf{1.0} &
  \multicolumn{1}{l|}{\textbf{0.14}} &
  \textbf{0.18} &
  \textbf{-3.15**} \\ \hline
Settings &
  12 &
  5 &
  \multicolumn{1}{l|}{0.98} &
  0.84* &
  \multicolumn{1}{l|}{0.82} &
  0.83 &
  \multicolumn{1}{l|}{0.8} &
  0.8 &
  \multicolumn{1}{l|}{0.13} &
  0.19 &
  -0.68 \\ \hline
\textbf{Fake Profiles} &
  \textbf{23} &
  \textbf{10} &
  \multicolumn{1}{l|}{\textbf{0.99}} &
  \textbf{0.98} &
  \multicolumn{1}{l|}{\textbf{0.49}} &
  \textbf{0.61} &
  \multicolumn{1}{l|}{\textbf{0.5}} &
  \textbf{0.6} &
  \multicolumn{1}{l|}{\textbf{0.15}} &
  \textbf{0.22} &
  \textbf{-3.08**} \\ \hline
\textbf{Social Tags} &
  \textbf{20} &
  \textbf{12} &
  \multicolumn{1}{l|}{\textbf{0.98}} &
  \textbf{0.96} &
  \multicolumn{1}{l|}{\textbf{0.45}} &
  \textbf{0.60} &
  \multicolumn{1}{l|}{\textbf{0.42}} &
  \textbf{0.67} &
  \multicolumn{1}{l|}{\textbf{0.22}} &
  \textbf{0.17} &
  \textbf{-4.35***} \\ \hline
\textbf{\begin{tabular}[c]{@{}l@{}}\edit{Safe}\\ Interactions\end{tabular}} &
  \textbf{14} &
  \textbf{9} &
  \multicolumn{1}{l|}{\textbf{0.77**}} &
  \textbf{0.99} &
  \multicolumn{1}{l|}{\textbf{0.67}} &
  \textbf{0.78} &
  \multicolumn{1}{l|}{\textbf{0.67}} &
  \textbf{0.78} &
  \multicolumn{1}{l|}{\textbf{0.14}} &
  \textbf{0.06} &
  \textbf{-3.07**} \\ \hline
\begin{tabular}[c]{@{}l@{}}Social Media\\ vs Reality\end{tabular} &
  13 &
  6 &
  \multicolumn{1}{l|}{0.96} &
  0.68*** &
  \multicolumn{1}{l|}{0.73} &
  0.86 &
  \multicolumn{1}{l|}{0.83} &
  1.0 &
  \multicolumn{1}{l|}{0.25} &
  0.22 &
  -1.84 \\ \hline
\end{tabular}
\caption{Pre=pre-intervention assessment; Post=post-intervention assessment; \textasciicircum{}Shapiro-Wilk test was used for determining normality; \textasciicircum{}\textasciicircum{}Means, Medians, and St.Dev are normalized by number of questions in each assessment; \textasciicircum{}\textasciicircum{}\textasciicircum{}Wilcoxon Signed Ranked test was used for assessing non-normal data; * p-values \textless{}= 0.05, ** \textless{}= 0.01; *** \textless{}= 0.001
}
\label{tab:quantresults}
\end{table}

In the following section, we report the quantitative differences between the pre- and post- assessments for students (shown in table ~\ref{tab:quantresults}). \oldedit{For the \edit{Platform Rules and Typical Use} module, we found there to be a statistically significant increase in test scores between the pre and post assessment} (\oldedit{n=25}, z=-3.15, p=0.0011, \edit{d=0.74}). The \edit{Platform Rules and Typical Use} module increased the median score from 0.8 to 1.0, with a mean difference of 0.11 and a 95\% confidence interval of -0.18 and -0.05. \oldedit{As the median post-test score of this module was 1.0, we may have seen a ceiling effect in participant scores. Despite this, we still saw a significant increase in participant scores. Our qualitative findings also demonstrate this. For instance, in response to being taught about the blocking functionality one participant shared, \textit{``Yeah, there's privacy settings on your posts, so you can say ``friends except for'' so you can block out who sees their posts'' (P12)}. This shows that the student was knowledgeable about the concept being taught. 

\oldedit{The Fake Profiles module also saw a statistically significant increase in score} (\oldedit{n=23}, t=-3.08, p=0.0102, \edit{d=0.51}). \oldedit{Average scores ranged} from 0.49 to 0.61, with a mean difference of 0.12, and a 95\% confidence interval of -0.20 and -0.04. \oldedit{Participants often compared the materials taught in this module to the decisions that they had made in the past: \textit{``Well, I actually did accept their friend request. But now looking at it... I answered 'no' for like almost all of [the criteria]. And so I ended up, like, rejecting it.'' (P7)}. This illustrates how participants could relate their learning in class to their past experiences to understand the concept.}}

\oldedit{In the Social Tags module, we saw a statistically significant increase in scores} (\oldedit{n=20}, t=-4.35, p=0.0003, \edit{d=1.08}). \oldedit{This module} led to an increased mean score from 0.45 to 0.61 with a mean difference of 0.16 and a 95\% confidence interval from -0.22 to -0.08. \oldedit{In this module, we had students express that they liked the definitions they were provided for social groups: \textit{``I like the definition... It's the middle one [pointing out on the slide], someone you've spoken only to on social media'' (P24)}. This indicates that they felt the definitions were applicable and made practical sense.} 

\oldedit{The \edit{Safe} Interactions module similarly had a significant increase in score} (\oldedit{n=14}, t=-3.07, p=0.0020, \edit{d=1.14}). \oldedit{For this module, the median score ranged} from 0.67 to 0.78 with a mean difference of 0.11 and a 95\% confidence interval from -0.17 to -0.05. \oldedit{While describing criteria for \finaledit{whom} it is safe to share certain types of information, students noted that they had a 'feeling' of whether or not certain types of information were safe\finaledit{:} \textit{``I had a feeling it wasn't safe to do that'' (P23)}. \finaledit{Similar} to the social tags module, this may indicate that students felt like the new rules aligned with their unspoken beliefs.} 

\oldedit{By contrast, the Settings module did not show significant changes in learning effect} (\oldedit{n=12}, z = -0.68 , p = 0.6562, \edit{d=0.07}). \oldedit{Our qualitative findings indicate that this may be because participants viewed this lesson as a tutorial of updating settings as opposed to material meant to be learned. For example, one student shared, \textit{``Do you want me to change it?'' (P10)} while researchers were summarizing the functionality of a setting.} 

\oldedit{Finally, the Social Media vs Reality module also did not show significant changes in learning effect} (\oldedit{n=13}, t = -1.84, p = 0.0977, \edit{d=0.73}). The median of the post-assessment for the Social Media vs Reality module was 1.0, indicating that there may be a ceiling effect with the results of this module. \oldedit{We saw this ceiling effect across participant responses. For instance, participants shared that they already knew the materials being taught\finaledit{:} \textit{``I already knew it was fake'' (P14)}.} \oldedit{Additionally,} it is notable that these two modules had the fewest number of participants. 

Of the lessons where participants showed significant improvement (\edit{Platform Rules and Typical Use}, Fake Profiles, Social Tags, \edit{Safe} Interactions), the change in mean ranged from 0.11 to 0.15 with an average mean change of 0.12. \edit{In \finaledit{other} words, the improvement was between 11\% and 15\% with the mean improvement being 12\%.} This indicates that students on average had a 12 percent increase in scores between the pre and post intervention. Among the 4 significant modules, a total of 41 questions (average:10.25) were asked. This is a meaningful increase as it shows that students got 1.23 more questions correct on average in the post-assessment than the pre-assessment.

\subsection{\oldedit{\edit{Classroom Engagement Patterns and Learning Outcomes} (RQ3)}}

\edit{In addition to evaluating learning gains (RQ2), we also examined how participants engaged during classroom sessions to better understand the relationship between engagement patterns and learning outcomes for autistic young adults. Prior research and neuro-affirming pedagogical perspectives suggest that autistic learners may engage in ways that differ from neuro-normative expectations of reciprocal or peer-directed interaction, particularly in group learning environments \cite{rotheram2010social, chen2021peer}. As such, RQ3 focuses on participants’ observable classroom engagement behaviors rather than traditional measures of collaboration in groups. Overall, participants rarely engaged in reciprocal communication with one another; instead, classroom engagement was primarily directed toward the instructor. Students asked and answered questions, made comments addressed to the instructor or the class as a whole, and shared tangential or loosely related information during discussions. The frequency of these engagement behaviors varied across students whose assessment scores improved, remained the same, or declined over the course of the intervention (Table~\ref{tab:rq3results}).}

\subsubsection{\edit{Answering Questions was Common Across All Students}}
Students from all outcome groups answered questions \edit{that were directly asked} in class. \edit{83.87\%} of students in the positive change group, \edit{77.78\%} of students in the no change group, and \edit{62.50\%} of students in the negative change group \edit{participated in this form of classroom engagement.} The answers which students provided to questions were also similarly detailed and correct regardless of group. For example, when participants were asked why they believe people create fake profiles, one participant in the negative change group shared, \textit{``It could be to get money or someone blocks you [and] you want to still contact them.'' (P12)}. A participant in the positive change group similarly shared, \textit{``Scam, cyberbully, be a predator, steal people's information. All that fun stuff'' (P6)}. \edit{This pattern suggests that when questions were asked explicitly and with sufficient structure, which is an approach aligned with autistic communication preferences \cite{wilson2021second}, students across outcome groups readily engaged, reflecting our intentional design of questions to support participation rather than to challenge students’ ability to respond.}

\begin{table}[] \edit{
\begin{tabular}{>{\raggedright\arraybackslash}p{0.1\linewidth}>{\raggedright\arraybackslash}p{0.12\linewidth}>{\raggedright\arraybackslash}p{0.12\linewidth}>{\raggedright\arraybackslash}p{0.12\linewidth}>{\raggedright\arraybackslash}p{0.12\linewidth}>{\raggedright\arraybackslash}p{0.12\linewidth}|>{\raggedright\arraybackslash}p{0.12\linewidth}}
\toprule
\textbf{ Pre/Post Change}&  \textbf{Answered} & \textbf{Asked- Class Content} & \textbf{Asked- Class Structure} 
 & \textbf{Share - Tangentially Related} 
 & \textbf{Share - Completely Unrelated}  
 & \textbf{No Verbal Classroom Engagement} \\
\midrule 

\multirow{2}{*}{\begin{tabular}[c]{@{}l@{}}
\textbf{Positive} \\
(N=31)
\end{tabular}}
& 83.87\% & 32.26\% & 22.58\% & 9.68\% & 16.13\% & 16.13\% \\
& (26 instances) & (10 instances) 
& (7 instances) & (3 instances) 
& (5 instances) & (5 instances) \\

\midrule

\multirow{2}{*}{\begin{tabular}[c]{@{}l@{}}
\textbf{No Change} \\
(N=9)
\end{tabular}}
& 77.78\% & 0.00\% & 0.00\% & 0.00\% & 0.00\% & 22.22\% \\
& (7 instances) & (0) & (0) & (0) & (0) 
& (2 instances) \\

\midrule

\multirow{2}{*}{\begin{tabular}[c]{@{}l@{}}
\textbf{Negative} \\
(N=8)
\end{tabular}}
& 75.00\% & 12.50\% & 0.00\% & 0.00\% & 0.00\% & 25.00\%\\
& (6 instances) & (1 instance) 
& (0) & (0) & (0) & (2 instances)\\

\bottomrule
\end{tabular} }
\caption{\edit{Percentage of Instances in Each Group where Students Would Participate in Various Behaviors}}
\label{tab:rq3results}
\end{table}
  
\subsubsection{\edit{Students with Improved Scores Were More Likely to Ask Questions}}
\edit{One of the main distinctions between groups \finaledit{was that} students who had a positive change in scores \finaledit{tended to ask} unprompted questions. There were a couple different types of questions which were asked by students. First, there were class content questions. These were questions that were directly related to the materials taught in class. For instance, \textcolor{black}{\textit{``Did you say that if they don't have a lot of information about them, then it's better to like reject?'' (P7)}}. 25.81\% (8 of 31) of students in the positive change group, 0\% (0 of 9) of students in the no change group, and 12.5\% (1 of 8) of students in the negative change group asked these types of questions. The other type of unprompted questions pertained to the class structure. These were questions where participants asked about what they should be doing in class. For example, \textit{``Are we not going to do the second activity?'' (P26)}. This behavior was also most common among students who had improved in score. 19.35\% (6 of 31) of the positive change group, 0\% (0 of 9) of the no change group, and 0\% (0 of 8) of the negative change group participated in asking class structure questions. This pattern suggest that unprompted question-asking, particularly questions seeking clarification of content or classroom structure, may reflect active sensemaking and comfort with the instructional environment and was more characteristic of students who demonstrated greater learning gains.}

\subsubsection{\edit{Only Students with Improved Scores Shared Tangentially Relevant or Completely Unrelated Information}}
\edit{The other main distinction between groups was sharing unrelated or tangential information.} These behaviors did not occur at all among the students we recorded in the no change or negative change group. They were prevalent in the positive change group. \edit{16.13\% (5 of 31) of those with positive change would share completely unrelated information during class. For example, \textit{``I went out with my girlfriend to Red Robin. I got the whiskey burger. Except then I didn't feel good after.'' (P4)}. 6.45\% (2 of 31) would share tangentially related information. This was information, that while unrelated to what was currently being taught in class, was still tangentially related to the topics in class. For example, one student shared, \textit{``'I love messing with people on scam calls' (P4)} when fake profiles were being discussed in class. Other times, students would comment on visual elements of class content: \textit{``Looks like box art for cereal or something''(P6).} 

Rather than indicating distraction or disengagement, these tangential contributions may reflect an alternative form of engagement and sensemaking that aligns with autistic communication styles. Prior work has shown that autistic learners often engage in \finaledit{ways that are} non-linear, associative, or less peer-reciprocal\finaledit{, but} nonetheless meaningful\finaledit{. N}euro-normative classroom expectations can misinterpret such engagement as off-task behavior rather than legitimate participation \cite{milton2012ontological, rotheram2010social}.}

\section{Discussion}
\oldedit{This paper is consistent with CSCW's focus on collaborative practices through PRISM, a classroom-based social media privacy intervention, co-designed with autistic young adults and educators. \edit{This is an important and novel area, as there is little information about teaching this user group about online safety, despite \finaledit{often being} targeted by malicious actors.} Our findings include empirical insights into how autistic YAs, a vulnerable population, perceive and navigate privacy risks on social platforms. Additionally, our results demonstrate that accessible, rule-based pedagogies, grounded in CSCW and design-based research\finaledit{,} can be effective in collaborative learning. \edit{This paper provides insights around how the PRISM curricula carefully supported learners to explore scenarios and share lived experiences of navigating the complexity of social media platforms and facing potential harmful content and actors.} This aligns with CSCW's emphasis on inclusion, accessibility, and equity.}

\subsection{\edit{Implications of Autistic Privacy Reasoning for Privacy Education (RQ1)}}

\begin{table}[b] \edit{
\centering
\caption{Core Privacy Beliefs and Strategies of Autistic Young Adults}
\label{tab:rq1beliefsactions_simplified}
\small
\begin{tabular}{p{0.32\linewidth} p{0.34\linewidth} p{0.32\linewidth}}
\toprule
\textbf{Belief / Strategy} & \textbf{Theme Summary} & \textbf{Related Autism Characteristics} \\
\midrule
\textbf{Anticipation of Social and Privacy Risk} &
Participants varied in how they anticipated online risk, including expectations of rejection, heightened concern about harm, or difficulty foreseeing risk. &
Differences in social risk perception \cite{gurbuz2024associations, lin2022autistic}; reliance on concrete meanings \cite{hobson2012autism}; double empathy problem \cite{milton2012ontological}. \\

\textbf{Categorical Reasoning About Safety} &
Privacy decisions were often based on absolute judgments about what is safe or unsafe rather than context. &
Preference for rule-based reasoning \cite{baron2009talent}; literal interpretation of safety guidance \cite{cullen2024towards}; reduced tolerance for ambiguity \cite{jenkinson2020relationship}. \\

\textbf{Rigid, Protective Privacy Strategies} &
Participants adopted all-or-nothing strategies such as blocking unknown users or limiting connections. &
Reliance on predictable strategies to manage uncertainty and reduce anxiety \cite{lung2024identifying}; reliance on environmental control \cite{gomot2012challenging, unwin2022use, lung2024identifying}. \\

\textbf{Avoidance as Self-Regulation} &
Platform avoidance or reduced participation was used to minimize unwanted interaction or perceived risk. &
Adaptive avoidance in response to social overload \cite{li2023autistic, lung2024identifying}; A need to maintain predictability \cite{gomot2012challenging}. \\

\textbf{Experience-Driven Rule Consolidation} &
Prior privacy harms led some participants to adopt inflexible rules to prevent recurrence. &
Self-regulatory coping strategies following negative experiences \cite{ng2022qualitative}; challenges with flexible generalization \cite{de2015brief}. \\
\bottomrule
\end{tabular} }
\end{table}

\edit{In RQ1, it was important to first understand autistic young adults’ existing privacy beliefs and management strategies before attempting to support adaptation to their mental models \cite{roberts2016review}. As summarized in Table~\ref{tab:rq1beliefsactions_simplified}, participants’ privacy beliefs and behaviors were closely intertwined with autism-related characteristics documented in prior literature, highlighting the importance of designing interventions that build from, rather than override, these existing strategies. Across participants, privacy beliefs frequently reflected how online risk was anticipated, including expectations of rejection, heightened concern about severe harm, or difficulty foreseeing risk altogether. Offline, anticipation of rejection among autistic individuals has been linked to early negative social experiences shaped by neuro-normative expectations \cite{lin2022autistic, gurbuz2024associations, lei2021have}. Our findings suggest that these experiences carry into online contexts, where some participants felt persistently vulnerable due to imagining catastrophic outcomes, while others expressed a sense of invulnerability by not anticipating risk at all. This polarization aligns with prior work on all-or-nothing thinking patterns in autism \cite{lage2024meta}. These findings suggest that privacy education should explicitly address polarized risk perception by helping learners reason through intermediate cases, rather than assuming risk is either obvious or nonexistent. Scenario-based instruction can provide structured opportunities to explore how risk varies across audiences, platform features, and contexts without relying on implicit social inference.

Participants’ reported privacy management strategies further reflected categorical reasoning and a desire to reduce uncertainty. As shown in Table~\ref{tab:rq1beliefsactions_simplified}, many participants relied on rigid, protective strategies such as blocking unknown users, limiting connections to established offline relationships, avoiding platforms altogether, or treating certain information as universally unsafe to share. These behaviors are consistent with autism-related tendencies toward reducing cognitive load, managing social ambiguity, and regulating exposure to potentially overwhelming environments \cite{lage2024meta, li2023autistic, bejerot2014social, zolyomi2019managing}. Rather than framing these strategies as maladaptive, our findings suggest that educational interventions should retain explicit, rule-based scaffolding while introducing nuance gradually, allowing learners to adapt rules across contexts without removing the predictability that makes them effective. Importantly, avoidance and disengagement emerged as common forms of self-regulation rather than lack of interest in social media. Participants’ use of blocking, platform avoidance, or limited participation functioned as protective mechanisms in response to perceived risk. This suggests that both educators and platform designers should treat avoidance as a signal of unmet support needs, and consider how clearer privacy controls, more legible affordances, and explicit guidance could enable safer engagement without requiring withdrawal from mainstream platforms.

Our findings also show that rigid rules did not always prevent harm. Some participants who relied heavily on inflexible privacy heuristics still reported negative online experiences, particularly after encountering scams or other violations. Consistent with prior work \cite{cullen2024towards, park2025teens}, these experience-driven rules often consolidated into absolute prohibitions that limited both risk and opportunity. This highlights the importance of privacy education that addresses not only how rules protect users, but also when rigid rules may fall short, supporting reflective reasoning about how safety, agency, and benefit can coexist. Overall, these findings support the need for privacy education that builds on autistic users’ strengths in rule-based reasoning while supporting contextual understanding. This directly validated the need for PRISM, which combines explicit decision rules with scenario-based context to empower autistic young adults to navigate mainstream social media platforms safely and on their own terms, without requiring conformity to neuro-normative interaction patterns or relegation to separate systems.}

\subsection{\edit{Evaluating the Efficacy of PRISM (RQ2)}}
\edit{To evaluate the efficacy of PRISM, we examined changes in participants’ social media privacy knowledge using pre- and post-assessments across the six instructional modules. Overall, the results indicate that PRISM was effective in improving privacy literacy for autistic young adults with level 2 support needs. In particular, modules that emphasized explicit, rule-based decision-making, such as identifying fake profiles and reasoning about social groups, demonstrated statistically significant learning gains. These findings align with prior research showing that rule-governed instruction can effectively support autonomy, safety, and decision-making for autistic individuals across contexts \cite{bradley2022rule, tarbox2011rule, cullen2024towards}. Beyond statistical significance, the pattern of results suggests that PRISM supported participants in moving beyond rigid, all-or-nothing mental models toward more adaptive privacy reasoning. Prior work has shown that autistic individuals often rely on categorical reasoning as a way to manage ambiguity and reduce cognitive load \cite{lage2024meta, mottron2006enhanced}. Our findings indicate that structured, rule-based educational scaffolding can serve as a productive starting point for introducing nuance, even when learners initially approach privacy decisions in binary terms. At the same time, social media privacy decisions are inherently contextual and may conflict with other social goals or values, underscoring that social nuance cannot always be captured through fixed rules alone \cite{akter2025calculating}.

Student engagement and learning outcomes were closely tied to content relevance and perceived utility. Modules grounded in participants’ lived experiences, such as managing online friendships, identifying unsafe profiles, or navigating social group boundaries, elicited stronger participation and reflection, consistent with prior work emphasizing the importance of personally meaningful learning goals in educational design \cite{dearden2017transforming}. In contrast, modules that were more procedural or abstract, such as changing platform settings, showed smaller improvements, suggesting that motivation and perceived relevance play a critical role in learning outcomes for this population. At the same time, high pre-assessment scores and ceiling effects observed in several modules indicate that some instructional content may have been too basic for our students. This finding highlights a wide-range of prior knowledge among autistic young adults and reinforces the importance of assuming competence when designing educational materials. Prior CSCW research has demonstrated the value of adaptive and personalized learning approaches, including nudging-based interventions \cite{wambsganss2022improving} and learner-sourced hints \cite{glassman2016learnersourcing}, which could be leveraged to dynamically tailor content difficulty, pacing, and scaffolding in future iterations of PRISM.

Finally, our findings suggest that hands-on or experiential learning should be incorporated thoughtfully. While prior work has shown that experiencing online harms can discourage future risky behavior among autistic individuals \cite{page2022perceiving}, relying on real-world negative outcomes as a primary educational mechanism risks exposing learners to unnecessary harm or triggering overly-protective responses. Instead, low-stakes, guided practice, such as scenario-based exploration or simulated decision-making, may offer a safer way to support learning without reproducing the consequences of privacy violations in the wild \cite{wisniewski2015preventative}. Such approaches can complement, rather than replace, structured instruction. Overall, these results support the efficacy of PRISM as a neuro-affirming, rule-based, and contextually grounded educational intervention for improving social media privacy literacy among autistic young adults. At the same time, they point to important opportunities for refinement, including greater personalization, careful calibration of content difficulty, and continued emphasis on empowering learners to make informed decisions aligned with their own goals and values.}

\subsection{\oldedit{The Effect of Participant Engagement in Learning Outcomes (RQ3)}}
\oldedit{We found that while all students would answer questions, only those who improved would ask questions or share unrelated or tangential content. This indicates that a key factor in \finaledit{student} performance is the level of self-censored engagement. Students who engage only in a way that is traditionally 'acceptable' in a classroom setting (answering questions correctly), but not in a way where they could come across as potentially vulnerable (asking questions) or off-task (sharing unrelated or tangential content) did not perform as well. Indeed prior work shows that higher levels of engagement are necessary to ask questions and lead to better learning outcomes \cite{chi2014icap}. \edit{Additionally, autistic communication styles are at times characterized by topic drift, or social interactions, that are perceived as off-task (i.e., "tangential comments). Some research has found that it occurs even more so for autistic individuals than the general population \cite{volden2002features}.} With the general population, the role of social interactions, which can be perceived as off-task, have been found to support collaborative problem-solving, building rapport, satisfaction, and learning outcomes \cite{langer2020exploring, huang2023social}. These findings also correlate with CSCW research which has found that informal chatting can motivate and promote social belonging for workers \cite{borghouts2022motivated, easley2023youths}. This may be especially important for Autistic individuals as these social supports have been shown to lead to improved learning outcomes \cite{holmes2024inclusion, cai2016educational, adreon2007evaluating}.} \edit{The notable absence of such behavior for the students who did not improve indicates that students may not have been fully at ease and exercising their typical communication styles. Future educational courses may need to watch for such cues as early indication of at-risk learners, and be very intentional in creating an environment that puts autistic learners at ease so they can succeed. }

\edit{ Neurotypical communication frequently assumes reciprocity and spontaneous verbal engagement, and deviations from these expectations may be negatively judged \cite{milton2012ontological}. Many autistic communicators, however, verbalize less when uncomfortable \cite{muris2021selective}, and being pressured into reciprocal interaction can increase distress or social anxiety, a common co-occurring condition \cite{bellini2006development, white2009anxiety}. These mismatched expectations contribute to the \textit{double empathy problem} in which autistic and neurotypical individuals struggle to interpret one another’s intentions \cite{milton2012ontological}. These findings challenge traditional CSCW notions of collaboration and group work, which often privilege reciprocal, peer-to-peer interaction as a marker of engagement and learning. In our classroom, collaboration rarely took the form of direct student-to-student exchange; instead, engagement was primarily instructor-directed, with students contributing through self-initiated questions or tangential commentary that was associated with learning gains. While such participation may be misinterpreted as off-task under neuro-normative expectations \cite{milton2012ontological}, our results suggest that CSCW conceptions of collaboration must be broadened to recognize non-reciprocal, non-linear, and individually mediated forms of engagement as legitimate and productive in learning environments designed for autistic participants.}

\subsection{Implications for Technology Educators and Social Media Designers}

\subsubsection{\edit{The Inclusive Design of Digital Literacy Programs for Autistic Users}}
\edit{We found that while our in-person course had significant and positive learning effects, students often chose only to interact with the teacher or aloud to the whole room, while person-to-person communication was limited. Work has found that autistic individuals may feel higher levels of distress when needing to engage in person-to-person communication \cite{bellini2010strength, white2009anxiety}. As such, we suggest that technology designers explore educational applications that support asynchronous participation and flexible interaction modalities (e.g., online), allowing autistic learners to engage with course content in ways that may be less socially demanding and more comfortable for them. Educational designers could incorporate features found beneficial for this population in prior work, such as opportunities to ask those in their close support network relevant questions, various forms of computer-mediated communication channels (e.g., chats, videos, forums), and recordings of class content \cite{hong2012designing, ringland2016will, das2021towards}. Based on our findings, digital literacy programs should offer multiple, optional pathways for participation, such as written or anonymous questions, instructor-directed interaction, and recorded content. Supporting these non-reciprocal and non-linear forms of engagement may better align with autistic learners’ preferences while still fostering meaningful learning outcomes.} 

\subsubsection{\edit{The Inclusive Design of Social Media Platforms for Autistic Users}}
\edit{Beyond educational interventions, our findings also have important implications for the design of mainstream social media platforms and the ways they support autistic users’ privacy decision-making. Prior work has found that social media can provide autistic users with important benefits \cite{wang2020benefits}. Despite this, mainstream social media platforms were designed to support neurotypical individuals rather than being inclusive of neurodiversity \cite{barros2023my}. While work from Hong et al., has examined building platforms specifically for those on the autism spectrum \cite{hong2012designing}, privacy was still a concern for a portion of participants. As our participants took an 'all-or-nothing' approach with using privacy settings (e.g., \textit{"block anyone you don't know"} (P12)), developing options that allow them to easily "lock-down" their social media privacy settings may support this. However, it is worth nothing that this may negate some of the positive effects of social media for this population, such as engaging in meaningful sociality through online communities \cite{ringland2016will}. This prioritizes the need for education, not just about the potential privacy harms, but also about how to navigate privacy decisions safely in order to still participate in online community-building. Based on our findings, we recommend that social media platforms offer graduated and transparent privacy controls that make intermediate choices easier to understand \cite{knijnenburg2022user}, rather than forcing users to choose all-or-nothing options. Platforms could also provide clearer explanations of why certain interactions or requests may be risky, supporting contextual reasoning without requiring users to rely solely on rigid rules. Finally, platforms should support low-risk pathways for participation, such as bounded community spaces or reversible privacy actions, to allow autistic users to explore social engagement while maintaining a sense of safety. These design implications reinforce the central argument of this paper that supporting autistic users’ privacy and participation on social media requires a combination of inclusive platform design and educational interventions that empower users to make informed decisions on their own terms.}

\subsection{Limitations and Future Research}
While this study offers important insights into how rule-based privacy education can support autistic young adults, several limitations warrant consideration and point to opportunities for future work. Our sample size was modest and, due to participant absences, not all students completed every pre- and post-assessment. The small class sizes—designed intentionally to accommodate students with substantial support needs—may have reduced statistical power, particularly for the Choosing Safer Privacy Settings and Social Media vs Reality modules. Future work should aim to replicate this intervention with a larger and more diverse cohort to strengthen statistical conclusions and increase generalizability. \finaledit{Additionally,  as participants were never required to speak, our ability to fully understand why participants may have decreased in score between the pre and post assessment was limited. From our work, it does appear that both students who went down in score and those who scored the same exhibited similar classroom behaviors. Future work can disentangle the differences between these groups.} Another limitation is the short-term nature of our assessment design. Pre- and post-tests were administered within the same session, providing a snapshot of immediate learning but not long-term retention or behavioral change. Although this structure aligned with prior findings on short-term memory challenges among autistic individuals \cite{desaunay2020memory, cheung2010verbal, southwick2011memory}, future research should incorporate follow-up assessments weeks or months after the intervention to better understand the durability and real-world impact of the learning gains observed.

Variability in assessment format, such as the number of questions and response types was another necessary trade-off. These differences helped reduce cognitive load and avoid participant fatigue, particularly given the wide range of support needs in our sample. While this decision limited direct comparisons across modules, our goal was to evaluate each module’s standalone effectiveness. Future iterations may explore more standardized yet accessible assessment tools. The lack of a control group also limits causal inference. However, given the ethical considerations of withholding potentially beneficial content and the logistical challenges of random assignment in a residential educational setting, we prioritized a within-subjects design. Future research might adopt quasi-experimental designs or delayed-intervention comparisons to enhance the strength of causal claims while remaining inclusive and ethically sound.

\edit{The educational modules were developed specifically for Facebook and Instagram. We chose to focus on these platforms based on there being prior work identifying the privacy risks of these platforms \cite{page2022perceiving} and recommendation from field site staff. As popular social technologies change, future work should look at expanding this educational intervention to focus on a variety of social media platforms.} Finally, we did not account for co-occurring conditions such as ADHD, which are common among autistic individuals \cite{stevens2016comorbidity}. Documentation challenges made it difficult to reliably identify comorbidities in our sample. Future studies could include ADHD screening tools such as the Adult ADHD Self-Report Scale (ASRS) \cite{kessler2005world} to better understand how overlapping cognitive profiles may shape learning needs and responses to rule-based interventions. Despite these limitations, this study demonstrates the feasibility and promise of tailoring privacy education to the cognitive and communication preferences of autistic young adults. By grounding curriculum design in both autism research and HCI principles, we offer a foundation for future interventions that can be further personalized, scaled, and extended to other aspects of digital safety and inclusion.

\section{Conclusion}
This study contributes to ongoing efforts in privacy education and inclusive HCI by developing and evaluating a classroom-based intervention tailored to the needs of autistic young adults with substantial support needs. By incorporating rule-based decision strategies and \finaledit{scenario-based formats} into the curriculum, the intervention aligns with cognitive styles commonly observed in this population and provides a foundation for more structured privacy decision-making. The introduction of the Privacy Rules for Inclusive Social Media (PRISM) framework offers a practical approach for teaching nuanced privacy concepts in a way that is accessible and developmentally appropriate. While preliminary, the results suggest that rule-based educational materials can support short-term improvements in privacy literacy. This work highlights the importance of adapting privacy education to diverse user needs and points to future opportunities for designing more inclusive tools and curricula that reflect the lived experiences of neurodiverse users.

\bibliographystyle{ACM-Reference-Format}
\bibliography{references}

\appendix



\section{Appendix A - Pre and Post Assessments}
\label{app:prepost}

\subsection{Platform Rules and Typical Use Pre-Post Assessment}
1. If you wanted to, would it be appropriate to have more than one account on Instagram? \\
\textbf{A. Yes} \\
B. No \\
C. I don't know \\

\noindent 2. You see George Clooney's (celebrity) account on Instagram. The account has a blue checkmark (figure ~\ref{fig:bluecheckmark}) next to it. What type of account is this?  \\
\begin{figure}[htp!]
    \centering
    \includegraphics[width=0.25\linewidth]{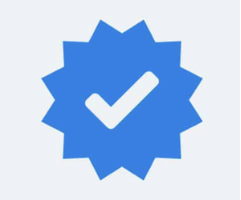}
    \caption{Blue Checkmark}
    \label{fig:bluecheckmark}
\end{figure}
\\ 
A. Personal account \\
B. Alternate personal account (finsta) \\ 
C. Business account \\
D. \textbf{Verified celebrity or influencer account} \\
E. Topic/interest account \\

\noindent 3. Your cousin created a second Instagram account just for their closest friends to follow. What type of account is this? \\
A. Personal account \\
B. \textbf{Alternate personal account (finsta)} \\ 
C. Business account \\
D. Verified celebrity or influencer account \\
E. Topic/interest account \\

\noindent 4. You see the Target (store) account on Instagram. What type of account is this? \\
A. Personal account \\
B. Alternate personal account (finsta) \\ 
C. \textbf{Business account} \\
D. Verified celebrity or influencer account \\
E. Topic/interest account \\

\noindent 5. You see an account on Instagram which only posts about Animal Crossing. What type of account is this? \\
A. Personal account \\
B. Alternate personal account (finsta) \\ 
C. Business account \\
D. Verified celebrity or influencer account \\
E. \textbf{Topic/interest account} \\

\noindent 6. You have an account on Instagram where you post photos of yourself and friends. What type of account is this? \\
A. \textbf{Personal account} \\
B. Alternate personal account (finsta) \\ 
C. Business account \\
D. Verified celebrity or influencer account \\
E. Topic/interest account \\

\noindent 7. If you wanted to, would it be appropriate to have have more than one account on Facebook? \\ 
A. Yes \\
\textbf{B. No} \\
C. I don't know \\

\noindent 8. (Fill in the blank) You can join a \_\_\_ where you can talk about your favorite video game with other Facebook users \\
A. Personal \\
B. Page \\
\textbf{C. Group} \\

\noindent 9. (Fill in the blank) Your mom sent you a friend request on Facebook. Her account is \_\_\_. \\
\textbf{A. Personal} \\
B. Page \\
C. Group \\

\noindent 10. (Fill in the blank) Your favorite actor has a \_\_\_ where they post photos from behind the scenes of their show.\\
A. Personal \\
\textbf{B. Page} \\
C. Group \\

\subsection{Settings Pre-Post Assessment}
\noindent 1. What does this setting do? (Figure ~\ref{fig:blocking}) \\
\begin{figure}[htp!]
    \centering
    \includegraphics[width=0.25\linewidth]{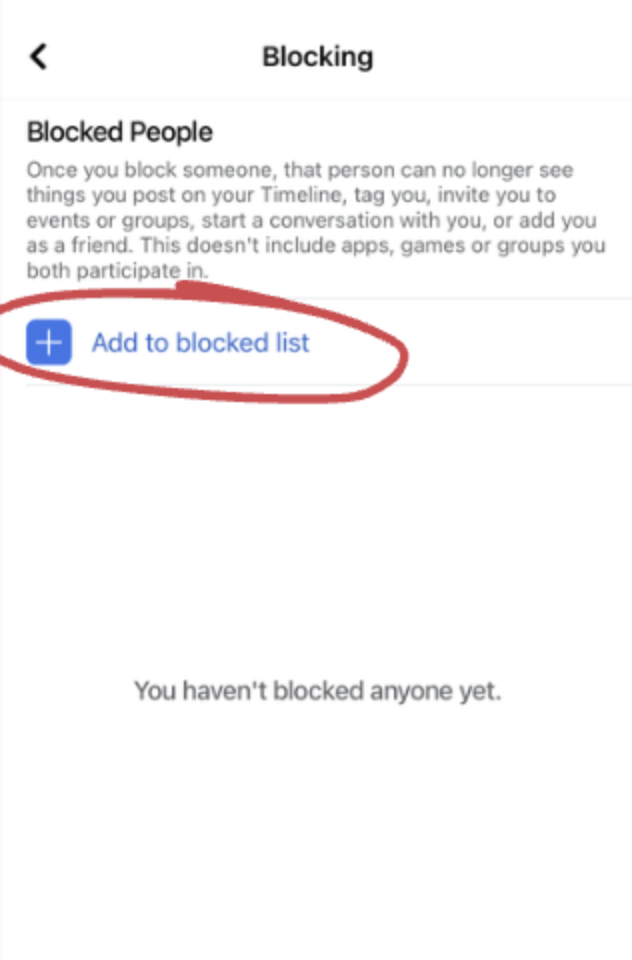}
    \caption{}
    \label{fig:blocking}
\end{figure}
\\ 
A. Updates who can tag you in their posts \\
B. Updates who can see your posts \\
C. Updates whether your account is private or public \\
\textbf{D. Allows you to block other users }\\
E. Updates where your messages go \\

\noindent 2. What does this setting do? (Figure ~\ref{fig:futureposts}) \\
\begin{figure}[htp!]
    \centering
    \includegraphics[width=0.25\linewidth]{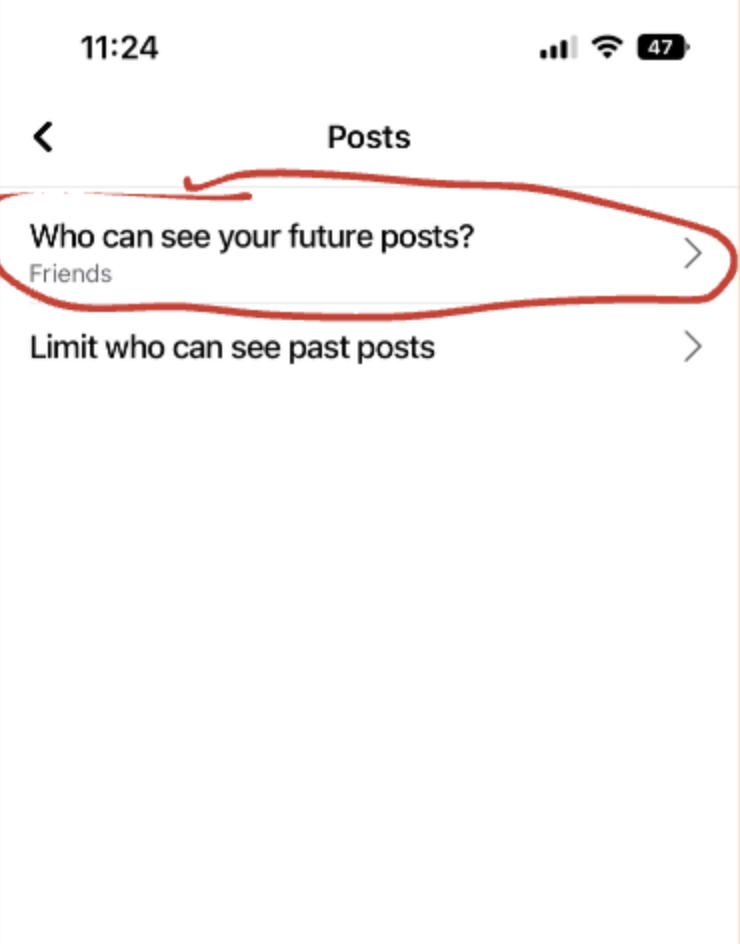}
    \caption{}
    \label{fig:futureposts}
\end{figure}
\\ 
A. Updates who can tag you in their posts \\
\textbf{B. Updates who can see your posts} \\
C. Updates whether your account is private or public \\
D. Allows you to block other users \\
E. Updates where your messages go \\

\noindent 3. What does this setting do? (Figure ~\ref{fig:private}) \\
\begin{figure}[htp!]
    \centering
    \includegraphics[width=0.25\linewidth]{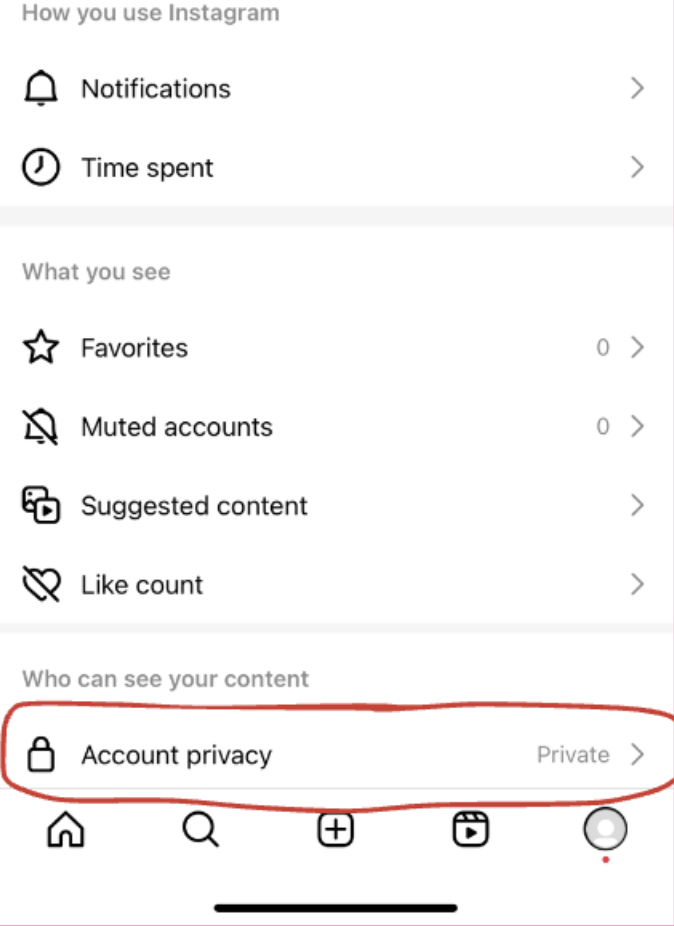}
    \caption{}
    \label{fig:private}
\end{figure}
\\ 
A. Updates who can tag you in their posts \\
B. Updates who can see your posts \\
\textbf{C. Updates whether your account is private or public} \\
D. Allows you to block other users \\
E. Updates where your messages go \\

\noindent 4. What does this setting do? (Figure ~\ref{fig:potentialconnections})\\
\begin{figure}[htp!]
    \centering
    \includegraphics[width=0.25\linewidth]{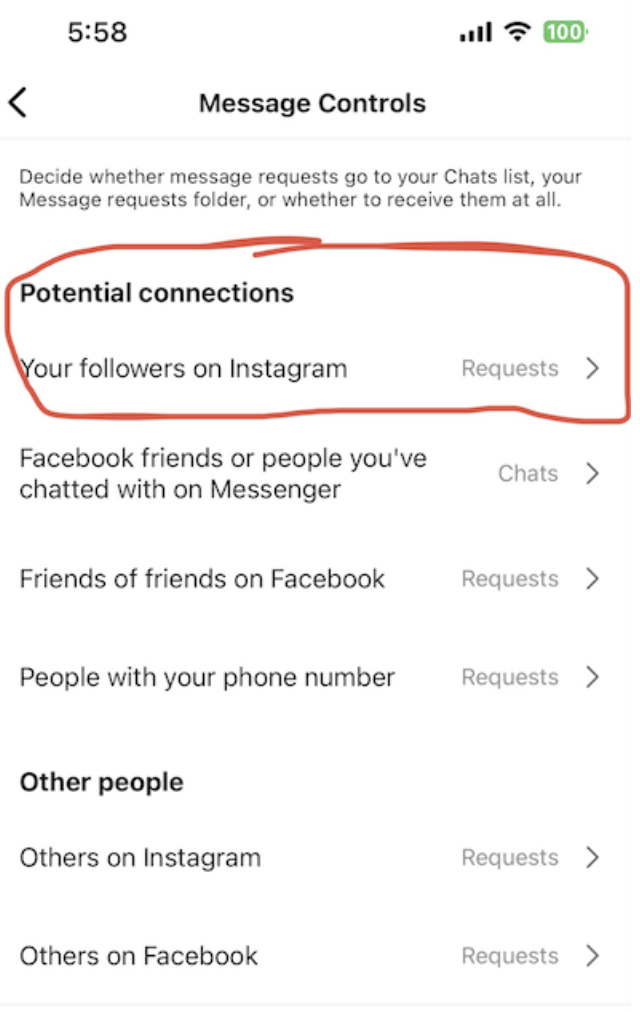}
    \caption{}
    \label{fig:potentialconnections}
\end{figure}
\\ 
A. Updates who can tag you in their posts \\
B. Updates who can see your posts \\
C. Updates whether your account is private or public \\
D. Allows you to block other users \\
\textbf{E. Updates where your messages go} \\

\noindent 5. What does this setting do? (Figure ~\ref{fig:tags})\\
\begin{figure}[htp!]
    \centering
    \includegraphics[width=0.25\linewidth]{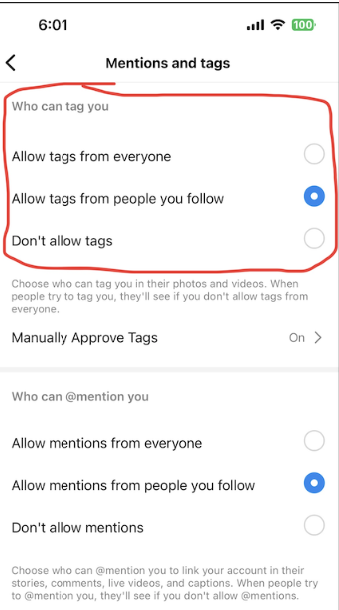}
    \caption{}
    \label{fig:tags}
\end{figure}
\\ 
\textbf{A. Updates who can tag you in their posts} \\
B. Updates who can see your posts \\
C. Updates whether your account is private or public \\
D. Allows you to block other users \\
E. Updates where your messages go \\

\subsection{Fake Profiles Pre-Post Assessment}
\begin{figure}[htp!]
    \centering
    \includegraphics[width=1\linewidth]{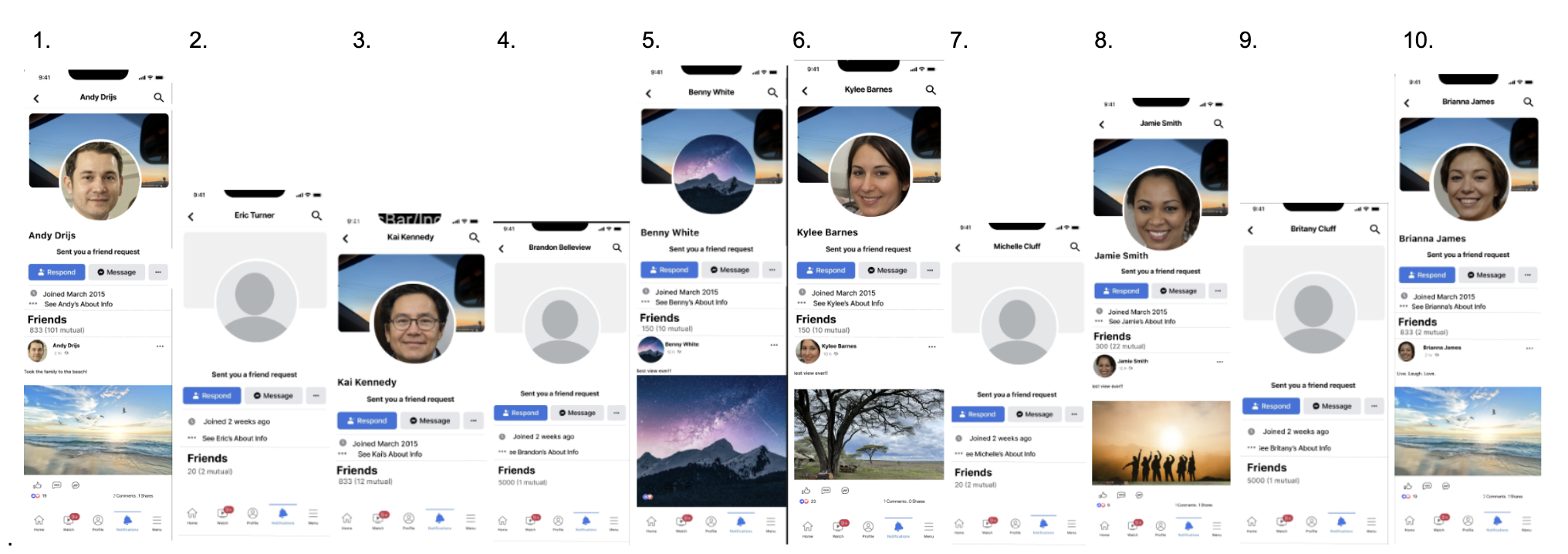}
    \caption{}
    \label{fig:fakeprofilesprofiles}
\end{figure}

\noindent 1. You are helping a friend decide whether or not to accept a friend request. They have met this person offline and the profile photo does match the person they know. This is the first connection request they have received from this person. (see profile 1 on figure ~\ref{fig:fakeprofilesprofiles})\\
Which of the following options should they do? \\
\textbf{A. Accept the request} \\
B. Reject the request \\
C. Ask the person who sent the request the next time they see them in-person \\
D. Ask an ally, family member, or mentor \\

\noindent 2. You are helping a friend decide whether or not to accept a friend request.
They have not met this person offline. This is not the first connection request they have received from this person. (see profile 2 on figure ~\ref{fig:fakeprofilesprofiles})\\
Which of the following options should they do? \\
A. Accept the request \\
\textbf{B. Reject the request} \\
C. Ask the person who sent the request the next time they see them in-person \\
D. Ask an ally, family member, or mentor \\

\noindent 3. You are helping a friend decide whether or not to accept a friend request.
They have not met this person offline. This is the first connection request they have received from this person. (see profile 3 figure ~\ref{fig:fakeprofilesprofiles})\\
Which of the following options should they do? \\
A. Accept the request \\
B. Reject the request \\
C. Ask the person who sent the request the next time they see them in-person \\
\textbf{D. Ask an ally, family member, or mentor} \\

\noindent 4. You are helping a friend decide whether or not to accept a friend request.
They have not met this person offline. This is the first connection request they have received from this person. (see profile 4 on figure ~\ref{fig:fakeprofilesprofiles})\\
Which of the following options should they do? \\
A. Accept the request \\
\textbf{B. Reject the request} \\
C. Ask the person who sent the request the next time they see them in-person \\
D. Ask an ally, family member, or mentor \\

\noindent 5. You are helping a friend decide whether or not to accept a friend request.
They have met this person offline. This is the first connection request they have received from this person. (see profile 5 on figure ~\ref{fig:fakeprofilesprofiles}) \\
Which of the following options should they do? \\
A. Accept the request \\
B. Reject the request \\
\textbf{C. Ask the person who sent the request the next time they see them in-person} \\
D. Ask an ally, family member, or mentor \\

\noindent 6. You are helping a friend decide whether or not to accept a friend request.
They have met this person offline and the profile photo does match the person they know. This is the first connection request they have received from this person. (see profile 6 on figure ~\ref{fig:fakeprofilesprofiles})\\
Which of the following options should they do? \\
\textbf{A. Accept the request} \\
B. Reject the request \\
C. Ask the person who sent the request the next time they see them in-person \\
D. Ask an ally, family member, or mentor \\

\noindent 7. You are helping a friend decide whether or not to accept a friend request.
They have not met this person offline. This is the first connection request they have received from this person. (see profile 7 on figure ~\ref{fig:fakeprofilesprofiles})\\
Which of the following options should they do? \\
A. Accept the request \\
\textbf{B. Reject the request} \\
C. Ask the person who sent the request the next time they see them in-person \\
D. Ask an ally, family member, or mentor \\

\noindent 8. You are helping a friend decide whether or not to accept a friend request.
They have not met this person offline. This is the first connection request they have received from this person. (see profile 8 on figure ~\ref{fig:fakeprofilesprofiles})\\
Which of the following options should they do? \\
A. Accept the request \\
B. Reject the request \\
C. Ask the person who sent the request the next time they see them in-person \\
\textbf{D. Ask an ally, family member, or mentor} \\

\noindent 9. You are helping a friend decide whether or not to accept a friend request.
They have met this person offline and the profile photo does not match the person they know. This is the first connection request they have received from this person.  (see profile 9 on figure ~\ref{fig:fakeprofilesprofiles})\\
Which of the following options should they do? \\
A. Accept the request \\
B. Reject the request \\
\textbf{C. Ask the person who sent the request the next time they see them in-person} \\
D. Ask an ally, family member, or mentor \\

\noindent 10. You are helping a friend decide whether or not to accept a friend request.
They have met this person offline and the profile photo does match the person they know. This is the first connection request they have received from this person. (see profile 10 on figure ~\ref{fig:fakeprofilesprofiles}) \\
Which of the following options should they do? \\
A. Accept the request \\
B. Reject the request \\
\textbf{C. Ask the person who sent the request the next time they see them in-person} \\
D. Ask an ally, family member, or mentor \\

\subsection{Social Tags}
\noindent 1. Kendra has been your neighbor for 10 years now and you enjoy spending time with her. For years, both of your families have done many things together. What social tag(s) apply to Kendra? \\
Choose all that apply \\
A. Family \\
\textbf{B. In-person friends} \\
C. School peers \\
D. Personal service providers \\
E. Work peers \\
F. Bosses and teachers \\
G. Online friends \\
H. Acquaintances \\
I. Community helpers \\
J. Strangers \\

\noindent 2. Amber is in your science class and she always seems to have the answers to the questions asked by your teacher. You never see Amber outside of class. What social tag(s) apply to Amber? \\
Choose all that apply \\
A. Family \\
B. In-person friends \\
\textbf{C. School peers} \\
D. Personal service providers \\
E. Work peers \\
F. Bosses and teachers \\
G. Online friends \\
H. Acquaintances \\
I. Community helpers \\
J. Strangers \\

\noindent 3. Andrew works with you at McDonalds. He taught you everything you know about flipping patties. While talking at work you discovered he loves playing chess just like you do! Ever since you learned about his passion, you’ve been playing chess and eating Wendy’s every weekend. What social tag(s) apply to Andrew? \\
Choose all that apply \\
A. Family \\
\textbf{B. In-person friends} \\
C. School peers \\
D. Personal service providers \\
\textbf{E. Work peers} \\
F. Bosses and teachers \\
G. Online friends \\
H. Acquaintances \\
I. Community helpers \\
J. Strangers \\

\noindent 4. Charlie and his family moved to your neighborhood 6 months ago. Last month you recognized Charlie at Church and decided to welcome him. You have since then seen him every Sunday but you don’t know much about him. What social tag(s) apply to Charlie? \\
Choose all that apply \\
A. Family \\
B. In-person friends \\
C. School peers \\
D. Personal service providers \\
E. Work peers \\
F. Bosses and teachers \\
G. Online friends \\
\textbf{H. Acquaintances} \\
I. Community helpers \\
J. Strangers \\

\noindent 5. Dr. Cole teaches your favorite math class. What social tag(s) apply to Dr. Cole? \\
Choose all that apply \\
A. Family \\
B. In-person friends \\
C. School peers \\
D. Personal service providers \\
E. Work peers \\
\textbf{F. Bosses and teachers} \\
G. Online friends \\
H. Acquaintances \\
I. Community helpers \\
J. Strangers \\

\noindent 6. Hannah is a job coach who you see weekly. You think Hannah is very attentive and always does her best to help you. What social tag(s) apply to Hannah? \\
Choose all that apply \\
A. Family \\
B. In-person friends \\
C. School peers \\
\textbf{D. Personal service providers} \\
E. Work peers \\
F. Bosses and teachers \\
G. Online friends \\
H. Acquaintances \\
I. Community helpers \\
J. Strangers \\

\noindent 7. Jenna got hired by your employer a year after you did. Your supervisor has asked you to train her. You've never seen her outside of work. What social tag(s) apply to Jenna. \\
Choose all that apply \\
A. Family \\
B. In-person friends \\
C. School peers \\
D. Personal service providers \\
\textbf{E. Work peers} \\
F. Bosses and teachers \\
G. Online friends \\
H. Acquaintances \\
I. Community helpers \\
J. Strangers \\

\noindent 8. Last week when grocery shopping, you dropped your shopping basket, dropping all of the fruits. A young man walking behind you with his daughter, saw what happened and helped you pick up your fruit. He said his name was Thomas. What social tag(s) apply to Thomas? \\
Choose all that apply \\
A. Family \\
B. In-person friends \\
C. School peers \\
D. Personal service providers \\
E. Work peers \\
F. Bosses and teachers \\
G. Online friends \\
H. Acquaintances \\
I. Community helpers \\
\textbf{J. Strangers} \\

\noindent 9. Michelle is taking the same algebra class as you do. You’ve both struggled on learning new concepts and thought it would be a good idea to work together on homework. You’ve both gotten really good at algebra and now hang out after school. What social tag(s) apply to Michelle? \\
Choose all that apply \\
A. Family \\
\textbf{B. In-person friends} \\
\textbf{C. School peers} \\
D. Personal service providers \\
E. Work peers \\
F. Bosses and teachers \\
G. Online friends \\
H. Acquaintances \\
I. Community helpers \\
J. Strangers \\

\noindent 10. You first met Eric while playing Minecraft. He was super good at it and you asked him to teach you how to build impressive things. While talking through the in-game feature you learned that Eric has your same age and you share the same interests. You now enjoy spending time while playing with Eric and even play other games together. What social tag(s) apply to Eric? \\
Choose all that apply \\
A. Family \\
B. In-person friends \\
C. School peers \\
D. Personal service providers \\
E. Work peers \\
F. Bosses and teachers \\
\textbf{G. Online friends} \\
H. Acquaintances \\
I. Community helpers \\
J. Strangers \\

\noindent 11. You love to spend time with your little brother Andrew. Andrew wants to spend more time with you, so he got a job at the same store where you work. What social tag(s) apply to Andrew? \\
Choose all that apply \\
\textbf{A. Family} \\
B. In-person friends \\
C. School peers \\
D. Personal service providers \\
\textbf{E. Work peers} \\
F. Bosses and teachers \\
G. Online friends \\
H. Acquaintances \\
I. Community helpers \\
J. Strangers \\

\noindent 12. Your Grandma May has signed up for a Facebook account and sent you a friend request. You remember that last year during Thanksgiving dinner, she mentioned something about wanting to communicate more often with her grandkids. What social tag(s) apply to Grandma May? \\
Choose all that apply \\
\textbf{A. Family} \\
B. In-person friends \\
C. School peers \\
D. Personal service providers \\
E. Work peers \\
F. Bosses and teachers \\
G. Online friends \\
H. Acquaintances \\
I. Community helpers \\
J. Strangers \\

\subsection{Safe Interactions Pre-Post Assessment}
\noindent 1. If you choose to, is it safe to give your personal email to a school peer over social media? \\
\textbf{A. Yes }\\
B. No \\

\noindent 2. If a school peer asks for your personal email over social media, is it safe to give it to them?  \\
A. Yes \\
\textbf{B. No} \\

\noindent 3. Is it safe to give your social security number to close family over social media?  \\
A. Yes \\
\textbf{B. No} \\

\noindent 4. Your online friend asks about the layout of your house, is it safe to give it to them?  \\
A. Yes \\
\textbf{B. No} \\

\noindent 5. A stranger online asked you to share your age range with them, is it safe to give it to them? \\
A. Yes \\
\textbf{B. No} \\

\noindent 6. An online friend asked you to share a picture of yourself at a concert with them, is it safe to send them this? \\
A. Yes \\
\textbf{B. No} \\

\noindent 7. You choose to share a picture of yourself at a concert with an online friend, is it safe to send them this?  \\
\textbf{A. Yes} \\
B. No \\

\noindent 8. A school peer asks you to share the layout of your home with them, is it safe to send them this?  \\
A. Yes \\
\textbf{B. No} \\

\noindent 9. You choose to share the layout of your home with a school peer, is it safe to send them this?  \\
A. Yes \\
\textbf{B. No} \\

\subsection{Social Media vs Reality}
\noindent 1. The pictures that I see on social media are always real \\
A. True \\
\textbf{B. False} \\

\noindent 2. If someone is smiling in a social media picture, that means they really are happy \\
A. True \\
\textbf{B. False} \\

\noindent 3. It's possible for someone to edit how they look in photos \\
\textbf{A. True} \\
B. False \\

\noindent 4. People can post false information on social media \\
\textbf{A. True} \\
B. False \\

\noindent 5. People always spread misinformation on purpose \\
A. True \\
\textbf{B. False} \\

\noindent 6. What should you do before re-posting news that you read on social media? (Select all that apply) \\
\textbf{A. Look it up} \\
\textbf{B. Be cautious} \\
C. Do nothing; automatically repost it \\
\textbf{D. Ask yourself about whether it seems real} \\

\end{document}